\documentclass[acmsmall, anonymous=false, balance]{acmart}

\usepackage{longtable}
\usepackage{multirow}
\usepackage{booktabs}
\usepackage[symbol]{footmisc}
\usepackage{graphicx}

\newcommand{\pogo}{\textit{Pok{\'e}mon GO}}
\newcommand{\hp}{\textit{Harry Potter: Wizards Unite}}
\newcommand{\ingress}{\textit{Ingress}}

\AtBeginDocument{%
  \providecommand\BibTeX{{%
    \normalfont B\kern-0.5em{\scshape i\kern-0.25em b}\kern-0.8em\TeX}}}

\setcopyright{acmcopyright}
\copyrightyear{2018}
\acmYear{2018}
\acmDOI{10.1145/1122445.1122456}

\acmJournal{JACM}
\acmVolume{37}
\acmNumber{4}
\acmArticle{111}
\acmMonth{8}

\begin{document}

\title[Casual and Hardcore Player Traits and Gratifications of Pok{\'e}mon GO, Harry Potter: Wizards Unite, Ingress]{Player Traits and Gratifications of Casual and Hardcore Players of Pok{\'e}mon GO, Harry Potter: Wizards Unite, and Ingress}

\author{John Dunham}
\affiliation{%
    \institution{Niantic x RIT Geo Games and Media Research Lab, Rochester Institute of Technology}
    \streetaddress{}
    \city{}
    \country{USA}}
    \email{jfd2017@rit.edu}

\author{Konstantinos Papangelis}
\affiliation{%
    \institution{Niantic x RIT Geo Games and Media Research Lab, Rochester Institute of Technology}
    \streetaddress{}
    \city{}
    \country{USA}}
    \email{kxpigm@rit.edu}
    
    \author{Nicolas LaLone}
\affiliation{%
    \institution{University of Nebraska Omaha}
    \streetaddress{}
    \city{}
    \country{USA}}
    \email{nlalone@unomaha.edu}
       \author{Yihong Wang}
\affiliation{%
    \institution{University of Liverpool}
    \streetaddress{}
    \city{}
    \country{UK}}
    \email{Yihong.Wang18@student.xjtlu.edu.cn}

\renewcommand{\shortauthors}{ et al.}

\begin{abstract}

Location-based games (LBG)  impose virtual spaces on top of physical locations. Studies have explored LBG from various perspectives. However, a comprehensive study of who these players are, their traits, their gratifications, and the links between them is conspicuously absent from the literature.  In this paper, we aim to address this lacuna through a series of surveys with 2390 active LBG players utilizing Tondello's Player Traits Model and Scale of Game playing Preferences, and Hamari's scale of LBG gratifications. Our findings (1) illustrate an association between player satisfaction and social aspects of the studied games, (2) explicate how the core-loops of the studied games impact the expressed gratifications and the affine traits of players, and (3) indicate a strong distinction between hardcore and casual players based on both traits and gratifications. Overall our findings shed light into the players of LBG, their traits, and gratifications they derive from playing LBGs.
\end{abstract}



\maketitle

\section{Introduction}
Location-based games (LBG) represent unique subset of video games. 
In the present research, we consider LBG a umbrella term for games which impose virtual space on the physical world.
These games incorporate mobile technology, and as alluded to above incorporate location as a central gameplay mechanic and are played in the real world. 
Since the advent of consumer grade GPS such games have been a part of the video game landscape, however LBG have risen to prominence with the release of Niantic's \pogo{} in 2016.
Today three games dominate the LBG market: \pogo{}, \hp{}, and \ingress{}. 
In an effort to better understand the players of LBG we examine these three games through the lens of player typologies and motivations.


While less prominent in LBG, player typologies have been the focus of much scholarly work in the context of video games.
The canonical “Hearts, Clubs, Diamonds, and Spades” typology, that was meant to typify the players of Multi-User Dungeons, served as a foundation to build upon with increasingly complex instrumentation \cite{bartle1996hearts}. 
Models such as the Game Experience Questionnaire (GEQ) \cite{ijsselsteijn2008measuring}, the Intensity, Sociability Games (InSoGa) heuristic model \cite{kallio2011least}, and the player satisfaction model called BrainHex \cite{nacke2014brainhex} have provided developers with instruments to assess their players and design experiences. 

Within many of these typologies, there is a binary or continuum of a colloquial category of hardcore and casual players that these instruments further typify. 
Traditionally, hardcore players are considered a grouping of players who favor more committed, higher intensity play. 
In contrast, lower intensity, less committed players are considered casual players. Whether presented as a binary  \cite{bateman2011player} or continuum  \cite{kallio2011least}, most typologies acknowledge the existence of this colloquial distinction. 
Typologies have been becoming increasingly sophisticated, leveraging more modern psychometric models with the intent of providing more accurate predictive modeling.
Recent work by Tondello has contextualized player typologies as a collection of traits  \cite{tondello2019dynamic}. 

Great strides have been made in general video gaming for player typologies, LBG research has, for the most, part eschewed this lens. 
The majority of research into LBG has been on their direct impacts on the players’ lives. 
For example, many studies have investigated the link between \pogo{} and physical health, with most finding positive links \cite{althoff2016influence,kaczmarek2017pikachu,kogan2017pilot}. 
In addition to the physical health of players, some studies have attempted to understand other potential health benefits and drawbacks of \pogo{} \cite{watanabe2017pokemon,lindqvist2018praise,wagner2017pokemon}. 
The exploration of player attributes and motivations is comparably lesser.  
Khalis and Mikami \cite{khalis2018s} have conducted exploratory work in assessing player traits of LBG players.  
While promising, the exploratory work in LBG player traits does not adequately advance trait research for traditional games. 

\pogo{} player motivations have recently been explored by Hamari  \cite{hamari2019uses}, albeit recontextualized as gratifications. 
Gratifications represent the player’s response to the dynamics of the games they play.
For example, the instrument would assess a player who enjoys social activities in a game as being gratified by socialization when surveyed for that game. 
In his work, Hamari  \cite{hamari2019uses} used gratifications in the context of LBG and their impact on in-app purchase intentions. 
As most LBG employ a free to play revenue model  \cite{liu2014effects}, such applications of gratifications are vital to commercial interests.

Our work builds and extends this line of work and synthesizes player traits, gratifications, and the hardcore/casual binary to better, more holistically understand the players of the studied LBG.
In detail, we explore the  following: (1)  The casual and hardcore players of \pogo{}, \hp{}, and \ingress{}, (2) the traits of casual and hardcore players of \pogo{}, \hp{}, and \ingress{}, and the (3) gratifications of casual and hardcore players of \pogo{}, \hp{}, and \ingress{}.

The paper is structured as follows. 
First, we describe the studied game’s core loops to contextualize the work. 
Next, we reviewed the current literature and state of the art on player typologies and LBG. 
We then describe the methodologies and data processing techniques employed in this study.  
This is followed by the results and by an in-depth discussion of said results.
Finally, we present the limitations of this work and avenues of future research.

\section{Differences Between \pogo{}, \hp{}, and \ingress{} based on their Core Gameplay Loops} \label{core-loops}


Prior to discussing the literature, it is important to contextualize the games we studied using the concept of core gameplay loops  \cite{tondello2017framework}. 
A core gameplay loop represents the cycle of interaction a player has with a game while playing it  \cite{sicart2015loops}.
In LBGs, the core gameplay loops generally reflect the relationship between a player's account, the physical space the player is in, amassing objects, and the concept of progression.
Each game's core gameplay loop will be detailed below.




In \ingress{}, the core gameplay loop can be distilled into four game dynamics: Explore, Collect Resources, Capture Portals, and Create Fields. 
Players begin the game with no resources and instructions to explore their surroundings. 
Exploration maps the virtual space of the game onto the physical reality occupied by the players, forcing the player to navigate in physical space for digital objectives. 
In the process of exploration, the player encounters resources which must be collected through the game’s interface. 
These resources are then used to capture portals for the player’s in-game team. 
When players on a team control portals, they may create fields between the portals to score points for their team. \ingress{} has a reasonably basic core loop and minimal polish (when compared to the other two games in this study); however, its dynamics appear designed to enforce team socialization and cooperation \cite{hatfield2014}.

\begin{figure}[htpb]
    \centering
    \includegraphics[width=.4\textwidth]{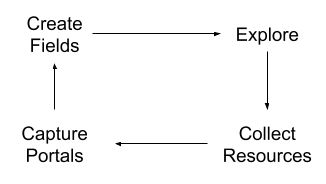}
    \caption{The Core Loop of \ingress{}}
    \label{fig:ingress-core-loop}
\end{figure}

The successor to \ingress{}, \pogo{} has an more complex core gameplay loop  \cite{fan2019}. 
As with \ingress{}, the game loop begins with the explore dynamic. 
Once again, exploration maps the virtual onto the physical, expecting the player to interact with the game through physical motion in the world. 
While exploring, the player encounters Pokémon, the player then captures Pokémon in these encounters. 
Players then gather resources from capturing additional Pokémon or visiting static locations in the world. 
Players level up their avatar and Pokémon, which then enables them to challenge gyms and raids. 
The dynamics which comprise \pogo{}’s core gameplay loop are more entwined than \ingress{}; however, the game focuses far more on the individual’s adventures in the virtual Pokémon world and less on social interactions and cooperation. 

\begin{figure}[htpb]
    \centering
    \includegraphics[width=.6\textwidth]{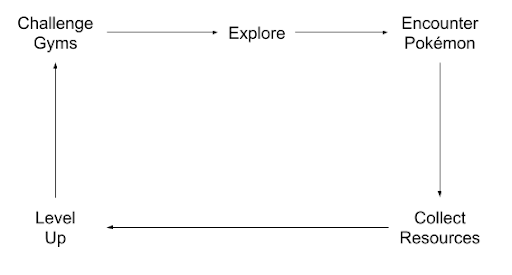}
    \caption{The Core Loop of \pogo{}  \cite{fan2019}}
    \label{fig:pokego-core-loop}
\end{figure}

While much more focused on the narrative related to the \textit{Harry Potter} franchise, the \hp{}’s core gameplay loop most closely resembles \pogo{}  \cite{kiiski2019}. 
The explore dynamic, similarly to \ingress{} and \pogo{}, acts as the entry point for the player into the game. 
A player is expected to gather ``foundables'', a dynamic resembling encountering/catching Pok{\'e}mon in \pogo{}. 
This dynamic directly fuels the complete foundable registry dynamic wherein players are encouraged to collect full sets of the foundable collectibles through a reward system. 
To get better at collecting foundables players must then gather currency and level up professions. 
Once players reach a sufficiently high level, they can play in co-op battles using the acquired skills. 
These co-op battles are more or less traditional dungeon crawls in which players may assemble a party with different skills to handle a diverse set of challenges earning foundables with more speed  \cite{kiiski2019}. 
This core loop is also entrenched in a more integrated story, however,  which may serve to focus the player more on their individual journey’s than the players of this game’s contemporaries.

\begin{figure}[htpb]
    \centering
    \includegraphics[width=.5\textwidth]{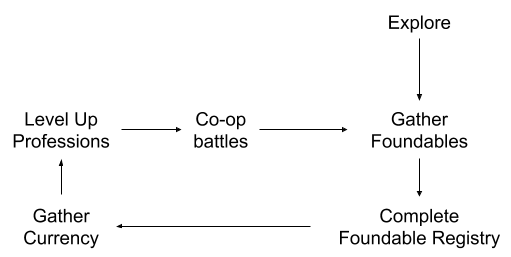}
    \caption{The Core Loop of \hp{}  \cite{kiiski2019}}
    \label{fig:HPWU-core-loop}
\end{figure}

In describing these games, it becomes fairly evident that \ingress{}, \pogo{}, and \hp{} have slightly different core loops. 
Despite those differences, the core loops are broadly similar and dynamic. 
The gameplay  revolves around an exploration mechanic realized through actually walking to physical locations and interacting briefly in-situ with gameplay elements.



\section{Background}
Extant research on games provides a framework for studying player traits and gratifications, and their relationship to the concept of casual and hardcore players in LBG. 
In this section, we outline a general history of the ways that players have been typographed, what the justification was for that particular typography, and its empirical consequence in both non-location-based and LBG. 
The often dichotomous representation of hardcore versus casual players is a continual presence in the study of players. 
Generally, hardcore has been defined as a grouping of more committed players with a higher intensity of play, and casual is typically defined a player grouping with low commitment to gaming. 
How and why this dichotomy appears, disappears, and is made into a continuum will also be explored here. 

Most histories of player typographies begin with Bartle’s typography for players of Multi-User Dungeons (MUDs) \cite{bartle1996hearts,bartle2005virtual}. 
In this theoretical model, two axes were defined, with the one axis representing social interactions in the game and the other interactions with the game’s world. 
On the game world axis, Achievers tend to seek rewards and points, while Explorers attempt to discover the secrets of the game world. 
On the social axis, Killers have a tendency to enjoy competitive gameplay, while Socializers attempt to form relationships and socialize with other players. 
Bartle \cite{bartle1996hearts,bartle2005virtual} would later modify this model by adding explicit and implicit motivations as a third axis. 
While useful for thinking about game design, this theoretical model was not empirically verified and no method for measuring the user’s dimensions was designed by Bartle \cite{yee2005motivations}.

In response to the lack of empirical grounding in Bartle’s model, Yee described an empirically-grounded model of player motivations \cite{yee2006motivations}.
Yee’s player motivations differ from Bartle’s typology teleologically. 
Where Bartle attempts to classify player personalities, Yee presents a possible explanation for why players play  \cite{yee2005motivations}. 
In this model there are three main components for describing player motivation: Achievement (Advancement, Mechanics, Competition), Social (Socializing, Relationship, Teamwork), and Immersion (Discovery, Role-Playing, Customization, Escapism). 
Players measured using this model are given component scores based on their answers to the model’s instrument. 
Each subcomponent was graded independently and offered a mapping of a player’s motivations  \cite{yee2005motivations}. 

Yee  \cite{yee2015gamer,yee2012online} would validate this model using data from Massively Multiplayer Online Role-playing games, specifically \textit{World of Warcraft} (N=2,716) and \textit{Everquest 2} (N=7,129)  \cite{williams2009looking}.
Later, this model would be expanded to  six clusters with twelve dimensions: Action (Destruction, Excitement), Social (Competition, Community), Mastery (Challenge, Strategy), Achievement (Completion, Power), Immersion (Fantasy, Story), and Creativity (Design, Discovery) \cite{yee2015gamer} . 
This iteration provides a more granular picture of the player’s motivations allowing designers to comprehend the dynamics of play preferred by players of their game. 
Yee overall did not specifically focus on the hardcore/casual binary; instead, these terms were distributed among the dimensions of Action, Mastery and Immersion. 
These terms are analogs for a continuum of hardcore and casual, with higher scores indicating more hardcore players. 

Contextually this was verified for MMOs and online games only but offered potential to understand player behavior in single-player or local multiplayer games. 
One such attempt at typographing players in other games was the Games Experience Questionnaire or GEQ. 
The GEQ was developed as “a self-report measure that aims to comprehensively and reliably characterize the multifaceted experience of playing digital games” for the fun of gaming (FUGA) project \cite{ijsselsteijn2008measuring}. 
At its core, the GEQ instrument is a modular questionnaire with multiple sub-components that allows researchers different types of fidelity of the player experience. 
These player experiences were categorized as post-game, social presence, and in-game. 
The instrument is to be used immediately after players completed gameplay sessions, with the intention of most closely capturing intention while playing the game. 
The GEQ also has a lighter module for data gathering during gameplay sessions. As is the case with Yee’s model, the GEQ doesn’t explicitly have a mechanism for gathering or observing the hardcore/casual dimension directly.

This instrument was deployed in multiple studies to analyze player experience. 
The types of research this instrument analyzed range from a study of the impact of Wiimotes on affective gameplay interactions (N=37) \cite{nacke2010wiimote} to an investigation on human visual attention \cite{vanden2015game}. 
While this research was well received by the academy, some felt that more instrumentation validation was needed. 
To wit, Law et al. carried out a literature review and validation study (N=633) of the GEQ \cite{law2018systematic}. 
The authors found inconsistent reporting of psychometric properties. 
The validation study failed to find evidence supporting the original GEQ study. 
As a result, the validity of research leveraging this, fairly popular, tool was essentially invalidated. The development of different approaches was needed.  

In contrast to the GEQ, Kallio presented a gamer mentalities model that heavily relies on the intensity of and sociability of the player’s gaming styles: committed gamers (frequent and/or long sessions), casual gamers (occasional and/or short play sessions), and gaming companions (play with their social group for accompaniment). 
These mentalities assume the hardcore/casual spectrum as a three position state: hardcore, casual and temporary player (a null set).

Kallio rejected this modelling of gamer mentalities during the statistical analysis of the initial survey and so re-tuned their instrumentation (N=4,000) \cite{kallio2011least}. 
The original hypothetical typology was found to be improperly descriptive due to the diverse nature of the surveyed gamer groups and overlap of suggested mentalities. 
It was found that sociability and intensity could take many forms so in place of the hypothetical model, a gaming heuristic was adopted in the form of the three-component Intensity, Sociability, Games (InSoGa) model \cite{kallio2011least}. 
The InSoGa model was developed through three qualitative studies that included 73 short interviews, 33 in-depth interviews, and 2 focus groups focused on digital gaming practices and experiences.

What Kaillo found was that the intensity of gaming is processed as a composition of three points: gaming session length, gaming regularity, and concentration  \cite{kallio2011least}. 
This is not a strictly linear relationship and intensity exists as a continuum of the three points. 
Next, the sociability of gaming is likewise a continuum combining the elements of physical space, virtual space, and metagaming. 
This particular dimension is relevant to the focus of this work, as LBG frequently appropriate physical space to meet playful ends \cite{hamari2019uses}. 
Finally, games consist of three indicators: individual games or devices, game genres, and accessibility. 
Kallio proposes that while players have the agency to choose the games they play, the games themselves have an inextricable impact on the mentality of the players. 
This sentiment is mirrored by Holm \cite{holm2019player}, albeit with a more granular focus on game mechanics \cite{kallio2011least}.

The components were then converted into three sets of nine total gaming mentalities.
Social mentalities, including Gaming with Children, Gaming with Mates, and Gaming for Company, outlined a collection of mentalities which “highlight the importance of accessibility, games and game devices”  \cite{bateman2011player}. 
These mentalities represented a less regular aspect of gaming, only occurring when players had the opportunity to socialize. 
This socialization exists outside of the casual/hardcore dichotomy and acts as more of a modifier to more hardcore and casual states. 
Casual mentalities in this model, Killing Time, Filling Gaps, and Relaxing, indicated mentalities unconcerned with commitment to gaming. 
In the InSoGa model a casual mentality is considered the opposite of committed. 
A casual player in this model tends to play games that they are familiar with, as a context switch, and/or to fill periods of time in their day. 
Gaming in the casual mentality is done when convenient. The hardcore representation in InSoGa model is represented by committed mentalities, Gaming for Fun, Immersive Play, and Gaming for Entertainment. 
In contrast with the casual mentalities, committed mentalities players find the act of playing itself as fun, lose themselves in the game, and/or game for gaming’s sake.

These mentalities are intended to be interpreted as heuristic representations of player motivations. 
Kallio found this lens to provide higher quality information about player motivation than strictly bounded typographies  \cite{kallio2011least}. 
The more nuanced approach to typographies does appear to allow for a more holistic understanding of players. 
Due to the lack of generalizability, there exists no simple method to measure this model and as a result, this work is best used as a framing device. 

In contrast to Kaillo’s qualitative measures of player mentalities, Bateman, Lowenhaupt, and Nacke \cite{bateman2011player} adapted the Myers-Briggs Type Indicator (MBTI) \cite{mccrae1989reinterpreting} in an effort to create an easily measured survey of player types. 
This instrument was called the Demographic Game Design mode. 
In this research, the hardcore and casual player classifications were assessed. 
Players were asked to self identify as hardcore or casual players. 
As Bateman, Lowenhaupt, and Nacke  \cite{bateman2011player} evaluated the data, this iteration of the model allowed them to identify four play styles: Conqueror, Manager, Wander, and Participant. 
These styles contrasted heavily with the existing Bartle types. 

Accordingly, this resulted in Bateman and Nacke  \cite{bateman2011player} asserting that hardcore and casual represented not a playstyle, but a trait dimension of the player’s personality. 
Bateman, Lowenhaupt, and Nacke \cite{bateman2011player} continued to posit that the concept of hardcore players may be better conveyed as gamer hobbyists, players who play a range, and diversity of many games.
In their view, a gamer hobbyist is a player who has a higher degree of game literacy, or a player’s ability to comprehend game mechanics, rules, and patterns in digital games  \cite{buckingham2007game} and generally has a capacity for more imaginative play. 
The promising results of DGD1 \cite{bateman200521st} led to further testing.
 
The DGD1 design \cite{bateman200521st} was iterated on incorporating the Temperament Theory described by Berens \cite{berens2000understanding}. 
Temperament theory suggests that there are four skill sets: Catalyst, Stabilizer, Strategic and Tactical, with each relating directly to a MBTI type result. 
The second Demographic Game Design Model, DGD2, re-purposed these skill sets into four hypothetical player archetypes: Logistical, Diplomatic, Strategic, and Tactical  \cite{bateman2011player}. 
When applied to the DGD2 survey (N=1,040) a bug prevented assessment of the Diplomatic skillset, but analysis found loading game literacy with Tactical skills, and Strategic with Logistical skills. 
The survey also identified that self selection for hardcore and casual may have issues when the binary is viewed through a game literacy lens, as respondents around 90\% of respondents claimed they had top marks in game literacy, despite only 50.1\% identifying as hardcore  \cite{bateman2011player}.

The DGD2 was immediately succeeded by the more systemic BrainHex \cite{nacke2014brainhex}.
This model was created from a collection of neurobiological papers and player surveys relating to the neurobiological basis of emotion. Hardcore versus casual players were not tracked by this study in lieu of more concrete neurobiological concepts.
The collection of player surveys suggested there are seven player types: Seeker (curiosity motivated), Survivor (fear motivated), Daredevil (excitement motivated), Mastermind (strategy motivated), Conqueror (challenge motivated), Socializer (social interaction motivated), and Achiever (goal motivated). 
Each of the seven player types were effectively models of motivation for players which could be used to foster deeper understanding of a game’s player base. 
In the introductory study, a survey was conducted (N=50,000) to establish relationships between personality types and the above mentioned player types \cite{nacke2014brainhex}. 

This typology moved more towards a motivation-based model, in line with Yee’s \cite{yee2015gamer} findings, moving away from the more rigid type model of Bartle \cite{bartle2005virtual}. 
BrainHex has since been used in research on tailoring persuasive health games to gamer types \cite{orji2013tailoring}, however, independent \cite{busch2016player,fortes2018towards} studies have found issues with its psychometric properties: the Seeker, Survivor, and Daredevil types required revision, while the Conqueror type may not have been distinct enough. 
Additionally, the MBTI is being replaced by trait theories in psychology literature \cite{mccrae1989reinterpreting}.
However, the authors note that BrainHex was a transitory model meant to foster more robust typologies or trait models \cite{bateman2011player}.

One model that grew out of the work on Brainhex was provided by Park \cite{park2011exploring} who proposed a link between personality traits and motivations to play online games. 
Park used the Big-Five  \cite{park2011exploring} personality domains (extraversion, agreeableness, conscientiousness, emotional stability and openness to experience) to evaluate the link between personality traits and motivations. 
Park performed an exploratory survey (N=524) measuring player demographics, motivations, genre preferences and the Big-Five personality domains using the Ten Item Personality Index, TIPI, instrument \cite{gosling2003very}.  
The TIPI is a collection of fifty questions, ten per domain, to generate an individual’s Big-Five personality scores. 
The instrumentation did not include affordances for the hardcore/casual binary.

Instead, Park discovered five motivations through factor analysis: Relationships, Adventure, Escapism, Relaxation, and Achievement  \cite{park2011exploring}. 
The central finding of the study was that agreeableness and extroversion acted as strong motivation predictors for playing online games; however, traits did not seem to be reliable predictors for player’s behavior.

The Big-Five personality domain and BrainHex were later combined by Tondello \cite{tondello2016gamification} as the gamification user types Hexad scale. 
The Hexad model’s user types are the intrinsic and extrinsic motivations defined in Self Determination Theory (SDT) \cite{ryan2000intrinsic}. 
As the name suggests the model features six user types each with their own motivations:

\begin{itemize}
    \item Philanthropists are motivated by purpose (altruism, no expectation of reward).
    \item Socializers are motivated by relatedness (interact with others, social connections).
    \item Free Spirits are motivated by autonomy (freedom to express themselves).
    \item Achievers are motivated by competence (seek to complete tasks, challenges).
    \item Players are motivated by extrinsic rewards (work to earn rewards in a system).
    \item Disruptors are motivated by triggering change (derived from empirical observation).
\end{itemize}

In the introduction of this model  \cite{tondello2016gamification}, Tondello carried out a survey (N=133) students at the University of Waterloo, with 40 agreeing to take a follow-up survey \cite{tondello2019don}. 
The survey consisted of the following: 

\begin{itemize}
    \item Demographic information: (age, gender, etc.)
    \item Hexad User Types: Thirty 7-point Likert scale questions in which participants were asked to assess how they aligned with statements (e.g. “I like to provoke” and “I like being part of a team”  \cite{tondello2019dynamic})
    \item Game element preferences: Thirty-two 7-point Likert scale questions asking participants about their affinity for specific game design elements (e.g. Easter Eggs, Quests, Customizing). These elements were pre-classified as being representative of certain types in the hexad scale.
    \item Personality: BIDR-6 \cite{winkler2006entwicklung} and BFI-10 \cite{rammstedt2007measuring} personality surveys were employed with a 7-point Likert scale.
\end{itemize}

The initial study was an effort to determine the efficacy of the Hexad model for modeling user game preferences, assess the scale’s correlation with participants’ personality traits  \cite{tondello2016gamification}, and verify the stability of the Hexad survey through a retest \cite{tondello2019don}. 
Tondello found the Hexad user types had positive correlations on all user types, except Philanthropist, with their expected design preferences  \cite{tondello2019don}. 
Additionally, the scale showed significant correlations with the Big-Five personality traits they had been theoretically designed. 
It was determined the instrument could be reduced to 24 entries with no loss in accuracy. 
Generally, the individual hexad types had the desired reliability in test and retest, however, the Player scale was determined to need additional work in future studies to improve reliability. 

The resulting 24 question instrument was validated and could theoretically be applied in the field with reasonable accuracy  \cite{tondello2019don}. 
However, it still needed refinement to address concerns stated above. 
Due to the overlap found with the Big-Five personality model  \cite{park2011exploring}, this type of model was further refined into a trait-based model by Tondello in further work. 
This trait model was proposed in an effort to “address the building blocks of actionable game design”  \cite{tondello2019don}. 

In pursuit of this goal Tondello conducted an exploratory survey of 350 participants, recruited through social media and mailing lists. 
Participants were incentivized to participate with the opportunity to win one of two \$50 international gift cards. 
Of the 350 participants, 332 responses were accepted after data cleaning and 157 agreed to take a follow up survey to calculate test-retest reliability. Of the 157 only 70  participants completed the retest  \cite{tondello2019don}.

This survey consisted of 50 items measured on a 7-point Likert scale, ten for each of the five proposed trait types:

\begin{itemize}
    \item Aesthetic orientation: high score correlates with players enjoying aesthetic experiences in games. Such experiences include world exploration or appreciation of the graphics, sound, and/or art style.
    \item Narrative orientation: high score corresponds to player enjoyment of complex narratives and stories within games.
    \item Goal orientation: players who score high in this trait enjoy completing goals in the game and tend to like 100\% completing games.
    \item Social orientation: high scorers typically enjoy playing with other players.
    \item Challenge (Action) orientation: high scoring players typically prefer games more difficult games and hard challenges. This trait more or less represents the hardcore/casual continuum in this model.
\end{itemize}

Initial factor analysis reduced the 50 items to 25, 5 for each of the traits. 
In the retest the reduced 25 item survey was deployed, and analysis showed stability of testing results despite the reduced number of items. 
Overall the findings of these surveys indicate a fairly reliable model of representing player traits. 
While distinct player traits are suspected to be tied to personality traits, it appears that personality is not the sole determinant of play preferences. 
Unlike prior typologies, as this trait model is a collection of five separate scores it’s less restrictive and allows for the full range of player preferences to be captured.

Models of player typologies and motivations are ubiquitous in the study of games.
However, this ubiquity is centered on more traditional video games.
The instruments provide researchers and designers a means to assess the desires and needs of player bases. 
Games such as \ingress{}, \pogo{}, and \hp{} have are somewhat different than traditional video games. 
When contextualized to LBG, these player typologies can help researchers and designers understand the players of these types of games. 

Kallio’s work \cite{kallio2011least}, as mentioned above, has clear applications in the LBG space. 
The rendering of the sociability indicator is easily adapted, with the physical space component represented by the act of players interacting with the world through the lens of a virtual space presented by the game itself. 
Further work has begun to take place in researching LBG, particularly in the context \pogo{}, and the impact of personality traits on player outcomes and behavior. 
Additionally, the existing modeling of hardcore/casual in the InSoGa provides researchers instrumentation to measure the hardcore/casual divide in the LBG space as well.

\pogo{} has been a routine target of inquiry since its release in July of 2016.
Preliminary research conducted near the release of the game by Althoff et.al. \cite{althoff2016influence} indicated  \pogo{} contributed significantly to increases in player physical activity. 
Exploratory research conducted by Kaczmarek et al. \cite{kaczmarek2017pikachu} in the form of an online survey (N=444) supported this finding. 
Players self reported that they had a greater incentive to spend more time outdoors and with greater amounts of physical activity. 
Kogan et al. likewise found that players were more likely to exercise with family and pets in an exploratory study (N=269) \cite{kogan2017pilot}. 
The benefits of LBG such as \pogo{} extend to mental health as well; a survey (N=2530) of Japanese workers by Watanabe et al. \cite{watanabe2017pokemon} found that players of the game had a significant improvement in their psychological distress over non-players.

LBG are not beyond critique, however, as several studies have outlined the inherent risks. 
Lindqvist et al. \cite{lindqvist2018praise} studied the effects of \pogo{} on eight families by collecting their experiences through focus groups. 
While generally positive, supporting the studies that indicated a positive correlation of increased physical activity with game play, Lindqvist et al. \cite{lindqvist2018praise} did outline potential safety concerns. 
In the study participants reported mild injuries which had occurred while playing the game (e.g. falling into a ditch), and parents expressed concern that the game would result in players not taking proper heed of their surroundings. 
A larger exploratory study (N=662) by Wagner-Green et al., supported this sentiment \cite{wagner2017pokemon}. 
Risky behaviors such as driving, biking and walking while focusing on the game were uncovered, and a third of players had reportedly lost sleep to play the game. 

Existing studies of LBG appear to support the conclusion that the benefits of such games (at least \pogo{}) are varied, albeit with some risk. 
Models of player traits in traditional gaming modalities focused on adjusting player outcomes and behaviors. 
It stands to reason that a deeper understanding of player traits, particularly in the context of LBG, might enable designers to encourage the more positive benefits of LBG. 
At the very least such work could attempt to discourage the aforementioned risky behavior. 
More in-depth work on player traits and motivations in LBG is currently taking form. 

In a step towards such research Khalis and Mikami  \cite{khalis2018s} explored player personality traits in the context of how they impacted player behaviors in \pogo{}. 
The exploratory study (N=101) used a questionnaire coupled with game data. 
It was found that higher social competence, conscientiousness, agreeableness, and lower social anxiety predicted the frequency in which players interacted with the game \cite{khalis2018s}. 
While not a strict survey of traits outlined in this background, these nascent findings outline the further study of trait theory on player behaviors in LBG.
	
In pursuit of improving understanding player motivation, Hamari \cite{hamari2019uses} conducted a study observing player gratifications and inhibitors. 
In the study, Hamari \cite{hamari2019uses} observed these gratifications and inhibitors and their impact on the player's intention to reuse (ITR) and in-app purchase intentions (IPI).
Methodologically, Hamari \cite{hamari2019uses} employed the use of a  player survey (N=1190) and modeled player gratifications across several dimensions assessed with use and gratifications. 
Challenge represents the player’s capabilities being tested against the game itself, Competition meanwhile represents a player’s capabilities being tested against other players. 
Both gratifications target similar dynamics, however, the key delimiter is their degree of socialization. 
When compared to Tondello’s \cite{tondello2016gamification} traits the sum total of these motivations represent the action trait.

Enjoyment represents the player's pleasure derived from playing the game, a gratification resembling the InSoGa \cite{kallio2011least} Gaming for Fun mentality. 
Trendiness is the social gratification of one’s gameplay being approved by peers. Socializing is the gratification received by interacting with other players of the game physically, a direct analog for Kaillo’s \cite{kallio2011least} social mentality. 
A unique attribute in digital games of LBG, Outdoor Activity, represented player gratification from playing outdoors. 
Due to the nature of \pogo{} as a target of childhood brand nostalgia, Hamari \cite{hamari2019uses} also measured Nostalgia, or a player’s gratification from a desire to relive their past. 
Ease of Use, Privacy Concerns, In-app Purchase Intentions and Intentions to Reuse were all also measured. 
Hamari \cite{hamari2019uses} found enjoyment, outdoor activity, ease of use, challenge, and nostalgia were all positively associated with ITR. 
Outdoor activity, challenge, competition, socializing, nostalgia and ITR were all associated with IPI. 
Privacy concerns and trendiness, meanwhile were found to have no association with ITR or IPI.

The results of this study have implications on the way LBG designers may design their game. 
It stands to reason that targeting dimensions that impact both ITR and IPI would result in the greatest yield for a studio attempting to drive in-app purchases.
For example, targeting ease of use would possibly have a greater impact than targeting trendiness or privacy concerns. 
While this study assessed the impact of gratifications on ITR and IPI, further research could re-target the effect being studied. Gratifications could be leveraged more specifically to determine what dimensions contribute positively to mental health or the strengthening of a community.	

Building on the state of the art this study attempts to get a clearer picture of the players of \pogo{}, \hp{}, and \ingress{}. 
A particular emphasis has been placed on studying the differences between the hardcore and casual players of these games. 
The models constructed by Tondello \cite{tondello2016gamification} and  Hamari \cite{hamari2019uses} have been leveraged as a framework to this end to explore the following about the players of \pogo{}, \hp{}, and \ingress{}:

\begin{enumerate}
    \item The casual and hardcore players.
    \item The traits of the casual and hardcore players.
    \item The gratifications casual and hardcore players derive from the games.
\end{enumerate}

To explore these topics, we evaluated the results of a survey instrument administered to a number of communities devoted to LBG.
Our collection of 2,390 survey responses represents the most in-depth evaluation of players of LBG to date.

\section{Data and Methods}

\subsection{Data Gathering and Demographics}

We recruited participants through the /r/PokemonGo, /r/HarryPotterWU, and/r/Ingress subreddits. 
Participants were required to be active players of one of either \pogo{}, \hp{}, or \ingress{} to participate. 
We deployed the questionnaires on mid-July and collected the responses for 10 days. 
During the period, there were in total 1253 participants in \pogo{} survey, 1281 participants in \hp{} survey and 310 participants in \ingress{} survey. 
However, we had to discard the responses that did not include answers to all the items in the questionnaires. 
In addition, responses which were observed to be sloppy (e.g. answer “agree” to all the items) were eliminated as well to ensure the quality of the dataset. 
Therefore, we have 1071, 1057 and 262 valid records remained in \pogo{}, \hp{} and \ingress{} respectively.
The demographic information regarding the gender, age, occupation, education and countries/regions of our participants are provided in Table \ref{survey-demographic-data}. 

\subsubsection{Gender} 
Most of the player in \pogo{} (74.4\%) and \ingress{} (80.5\%) are males. 
Notably, there is a gender balance in \hp{} as 51.1\% of the participants are males while 47.1\% of the participants are females. 
The percentage of non-binary payers in all the 3 games is low ranging from 1.5\% to 2.1\%. 

\subsubsection{Age} 
Approximately two thirds of the players in \pogo{} (65.3\%) and \hp{} (68.4\%) are young players between 21 and 35 years old while elder players show more interests in \ingress{} as half of the players in \ingress{} are 36 years old or above. 
In addition, \pogo{} is more popular among kids and teenagers given around one fifth of the participants (20.4\%) in this game are under 20 years old while only 3.9\% and 7.6\% of the players in \hp{} and \ingress{} belong to the same age group. 

\subsubsection{Occupation} 
Most of the players in \pogo{} (60.0\%), \hp{} (76.9\%) and \ingress{} (76.7\%) are employed while the second largest group of players in all the 3 games are students.
In comparison to \hp{} and \ingress{} in which students account for slightly over 10\% of the total sample, nearly 1/3 of participants in \pogo{} are students.
This is in accordance with the finding in Age which illustrates that \pogo{} is more popular among young people.

\subsubsection{Education} 
Over half of the participants in \pogo{} (50.6\%), \hp{} (64.4\%) and \ingress{} (50.8\%) reported that they graduated from University with bachelor’s degrees or above. 
Moreover, approximately one fifths of the participants in all 3 games have college or vocational degrees. This indicates that most of the players in our survey are well-educated. In addition, it is unsurprising to find the 33.2\% of the participants in \pogo{} have high school or lower degrees since many players are found to be students at less than 20 years old. 
However, it is interesting to find that similar percentage of participants (31.7\%) in \ingress{} also have the relatively low education level with high school or lower degrees given half of the players are 36 years or elder. 

\subsubsection{Countries/Regions} 
LBG Players from over 60 countries/regions around the world responded to our survey. 
However, in order to keep the data clear and informative, we decide to report the top 10 countries according the number of respondents. 
Based on the statistics, most of the participants in \pogo{} (66.9\%), \hp{} (66.9\%) and \ingress{} (70.2\%) are from English-speaking countries namely United States, United Kingdom, Canada and Australia. 
In addition, European countries such as Germany, Netherlands, Poland and Finland also contributed a considerable number of participants in our surveys. 

\begin{longtable}{|l|l|l|l|}
\hline
\multicolumn{1}{|c|}{\textbf{Characteristics}} & \textbf{\pogo{} (N=1071)} & \textbf{HPWU (N=1057)} & \textbf{\ingress{} (N=262)} \\ \hline
\endfirsthead
\endhead
\multicolumn{4}{|c|}{\textbf{Gender}} \\ \hline
\textbf{Male} & 797 (74.4\%) & 540 (51.1\%) & 211 (80.5\%) \\ \hline
\textbf{Female} & 252 (23.5\%) & 498 (47.1\%) & 47 (17.9\%) \\ \hline
\textbf{Non-binary} & 22 (2.1\%) & 19 (1.8\%) & 4 (1.5\%) \\ \hline
\multicolumn{4}{|c|}{\textbf{Age}} \\ \hline
\textbf{< 21 years} & 219 (20.4\%) & 41 (3.9\%) & 20 (7.6\%) \\ \hline
\textbf{21-35 years} & 699 (65.3\%) & 723 (68.4\%) & 98 (37.4\%) \\ \hline
\textbf{> 35 years} & 153 (14.3\%) & 293 (27.7\%) & 144 (50\%) \\ \hline
\multicolumn{4}{|c|}{\textbf{Occupation}} \\ \hline
\textbf{Employed} & 642 (60.0\%) & 813 (76.9\%) & 201 (76.7\%) \\ \hline
\textbf{Unemployed} & 95 (8.9\%) & 94 (8.9\%) & 29 (11.1\%) \\ \hline
\textbf{Student} & 334 (31.1 \%) & 150 (14.2\%) & 32 (12.2\%) \\ \hline
\multicolumn{4}{|c|}{\textbf{Education}} \\ \hline
\textbf{University degree or above} & 542 (50.6\%) & 681 (64.4\%) & 133 (50.8\%) \\ \hline
\textbf{College degree} & 129 (12.1\%) & 173 (16.4\%) & 41 (15.6\%) \\ \hline
\textbf{Vocational degree} & 44 (4.1\%) & 34 (3.2\%) & 5 (1.9\%) \\ \hline
\textbf{High school or lower} & 356 (33.2\%) & 169 (16.0\%) & 83 (31.7\%) \\ \hline
\multicolumn{4}{|c|}{\textbf{Countries/Regions}} \\ \hline
\textbf{United States} & 498 (46.5\%) & 534 (50.5\%) & 126 (48.1\%) \\ \hline
\textbf{United Kingdom} & 123 (11.5\%) & 93 (8.8\%) & 44 (16.8\%) \\ \hline
\textbf{Germany} & 65 (6.1\%) & 56 (5.3\%) & 17 (6.5\%) \\ \hline
\textbf{Canada} & 56 (5.2\%) & 60 (5.7\%) & 11 (4.2\%) \\ \hline
\textbf{Australia} & 40 (3.7\%) & 20 (1.9\%) & 3 (1.1\%) \\ \hline
\textbf{Netherlands} & 24 (2.2\%) & 25 (2.4\%) & 5 (1.9\%) \\ \hline
\textbf{Poland} & 19 (1.8\%) & 14 (1.3\%) & 1 (0.4\%) \\ \hline
\textbf{Italy} & 17 (1.6\%) & 12 (1.1\%) & 1 (0.4\%) \\ \hline
\textbf{France} & 16 (1.5\%) & 13 (1.2\%) & 1 (0.4\%) \\ \hline
\textbf{Finland} & 15 (1.4\%) & 17 (1.6\%) & 4 (1.5\%) \\ \hline
\textbf{Others} & 198 (18.4\%) & 210 (19.9\%) & 49 (18.7\%) \\ \hline
\caption{Demographic Data from our Survey}
\label{survey-demographic-data}
\end{longtable}

\subsection{Data Preparation and Cleaning}

\subsubsection{Data Filtering} \label{data-filter}

Before analyzing the validity and reliability of our questionnaires, we initially eliminated the records with missing data as well as the outliers detected through our observation and SPSS built-in functions. 
This is based on the theories that (1) the missing data and outliers should be removed in order to ensure the quality of the data, and (2) that missing data and outliers would have serious impacts on the analysis of EFA  \cite{liu2012demonstration}. 
Therefore, there are 1071, 1057 and 262 valid records remained in \pogo{}, \hp{} and \ingress{} respectively.

In addition, based on the suggestions from Tabachnick and Fidell  \cite{tabachnick2007using} we conducted normality test on the sample in \ingress{} through checking the skewness and kurtosis and found that there are no items involved in the questionnaires were beyond the recommended thresholds. 
The reasons why we did not conduct normality test on samples in \pogo{} and \hp{} were because that previous study suggested that in large dataset most statistical methods rely on the Central Limit Theorem, which states that the average of a large number of independent random variables is approximately normally distributed around the true population mean  \cite{lumley2002importance}. 
Furthermore, in large sample the violation of normality should not cause major problems and one should not even worry about the normality in a dataset consisting of over 200 participants  \cite{elliott2007statistical,fielddiscovering}.
Therefore, all the items were retained for further analysis of the questionnaires in this study. 

\subsection{Instrument Reliability and Validity}

\subsubsection{Validity Test} \label{CFA-EFA}
After readying the data, we decided to conduct both exploratory factor analysis (EFA) and confirmatory factor analysis (CFA) to test the validity of the questionnaires we used in this study. Although one could test only CFA if there are solid assumptions about the structure of the factors being tested \cite{henson2006use}, we found that it may be risky to skip the EFA in our case. 
This is because that (1) both the Traits questionnaire and the Gratification questionnaire we used were newly published with limited empirical researches grounded on them; (2) data from different samples may result in different factor solution of the measuring instruments \cite{gaskin2017sample}. 
Therefore, based on the suggestions from Hair et al.  \cite{hair1998multivariate}, we split the sample in each survey in half and conduct EFA on the first half of the data to extract the underlying factor structures in the questionnaires and, then, do the CFA on the other half of the data to confirm the identified structures are valid.

Exploratory factor analysis (EFA) was conducted using SPSS. First, simple principal components analysis was conducted. 
It was recommended by Kaiser \cite{kaiser1960application} that the eigenvalue of the extracted factor be higher than 1.00. 
Secondly, a principal component analysis with a varimax rotation was studied to analyze the possible solutions. 
According to \cite{fielddiscovering,decoster1998overview,hair1998multivariate}, the following criteria were set to determine if the dataset is suitable for EFA and which items should be retained in the test:
 
\begin{itemize}
    \item Kaiser–Meyer-Olkin (KMO) should be above 0.6 
    \item Bartlett’s Test of Sphericity should be significant at $\alpha$ < .05
    \item The communalities of items should be above 0.3
    \item Loading of item should be more than 0.32. 
    \item An item should not be permitted to load on two or more factors.
    \item Retained items and factors should make sense conceptually.
    \item A factor in the model of principal component analysis should have at least 3 items.
\end{itemize}

To test the factorial validity of Traits and Gratification sections in our surveys, confirmatory factor analysis (CFA) was conducted using the structural equation modelling (SEM) method with the help of the AMOS program. The following indexes were examined and reported in this study according to the commonly accepted standards  \cite{bennett2002comparison,kelloway2014using,hair1998multivariate,hancock2013structural,yan2019chinese}:

\begin{longtable}[htpb]{|c|l|}
\hline
\textbf{Index} & \textbf{Criteria} \\ \hline
\endfirsthead
\endhead
\multirow{2}{*}{\textbf{chi square to degree of freedom ratio ($\chi$2/df)}} & Good: $\chi$2/df \textless 3 \\ \cline{2-2} 
 & Acceptable: $\chi$2/df \textless{}5 \\ \hline
\multirow{2}{*}{\textbf{Standardized Root Mean Square Residual (SRMR)}} & Good: SRMR\textless{}0.08 \\ \cline{2-2} 
 & Acceptable: SRMR \textless{}0.1 \\ \hline
\multirow{2}{*}{\textbf{Root Mean Square Error of Approximation (RMSEA)}} & Good: RMSEA \textless{}0.08 \\ \cline{2-2} 
 & Acceptable: RMSEA\textless{}0.1 \\ \hline
\multirow{2}{*}{\textbf{Tucker Lewis index (TLI)}} & Good: TLI\textgreater{}0.9 \\ \cline{2-2} 
 & Acceptable: TLI\textgreater{}0.8 \\ \hline
\multirow{2}{*}{\textbf{Comparative fit index (CFI)}} & Good: CFI\textgreater{}0.9 \\ \cline{2-2} 
 & Acceptable: CFI\textgreater{}0.8 \\ \hline
\caption{These are the typical criterion for good and acceptable statistical results}
\label{what-is-good-table}
\end{longtable}

\subsubsection{Reliability test}

Cronbach’s alpha was used to investigate the reliability of obtained factors. 
Although 0.70 is usually considered as an acceptable cut-off point \cite{nunnally1994psychometric}, it is argued by Hair et al. \cite{hair1998multivariate} that a Cronbach’s alpha of more than 0.6 indicates a reasonable internal consistency. 
Therefore, the lowest standard of Cronbach’s alpha in this study was set to be at least 0.6.

In order to investigate the traits of casual and hardcore players of \pogo{}, \hp{}, and \ingress{}, we utilized a traits model and scale of game playing preferences that is proposed by Tondello  \cite{tondello2019don}. 
This measurement consists of 25 items in 5 dimensions, namely (1) Social orientation; (2) Action (Challenge) orientation; (3) Immersion (Narrative) orientation; (4) Aesthetic orientation and (5) Goal orientation.

\subsection{Reliability and Validity of Traits Section of Survey}

\subsubsection{\pogo{}}

\paragraph{Results of EFA}

The results KMO and Barlett’s test indicate that the data in Traits section is suitable for exploratory factor analysis, given a KMO > 0.6 suggests that the sample is large enough to perform EFA and the significance in Barlett’s test (Sig. < 0.05) indicates that the correlation between items were sufficiently large.

\begin{longtable}[htpb]{|l|l|l|}
\hline
\multicolumn{3}{|c|}{\textbf{KMO and Bartlett's Test}} \\ \hline
\endfirsthead
\endhead
\multicolumn{2}{|l|}{Kaiser-Meyer-Olkin Measure of Sampling Adequacy} & \textbf{.869} \\ \hline
\multirow{3}{*}{Bartlett's Test of Sphericity} & Approx. Chi-Square & 8236.408 \\ \cline{2-3} 
 & df & 300 \\ \cline{2-3} 
 & Sig. & \textbf{.000} \\ \hline
\caption{Results from the KMO and Bartlet's Tests for our \pogo{} Survey} 
\label{pokego-kmo}
\end{longtable}

Based on the aforementioned criteria regarding the retention of items (see section \ref{data-filter}), no items were removed in the exploratory factor analysis of the Traits section in \pogo{} survey. 
The factor loadings and communalities as well as the extracted factors and their cumulative explanations of the variance are shown in the following table. 
The results illustrated that all items correspond perfectly to the dimensions described in Tondello’s paper. 
Therefore, the first factor is “Social orientation”, which states the player’s preference for playing together with others online or in the same space. 
Factor 2 is “Action (Challenge) orientation”, which describes the player’s preference for challenging and fast-paced gameplay. 
The third factor is “Immersion (Narrative) orientation”, which reveals the player’s preference for complex stories or narratives within the games. 
Factor 4 is “Goal orientation”, which is related to player’s preference for gameplay that involves completing quests or tasks. 
The last factor is about the player’s preference for aesthetic experiences such as exploring the game world and appreciating the sound, graphics and art style in the game. 
Therefore, it is named by Tondello as “Aesthetic orientation”. 68.316\% of the variance was explained by the 5 factors in total.

\begin{table}[htpb]
\resizebox{\textwidth}{!}{%
\begin{tabular}{|p{6.5cm}|l|l|l|l|l|l|}
\hline
\multicolumn{1}{|c|}{\multirow{2}{*}{\textbf{Item}}} & \multicolumn{5}{c|}{\textbf{Component}} & \multirow{2}{*}{\textbf{Communality}} \\ \cline{2-6}
\multicolumn{1}{|c|}{} & \textbf{1} & \textbf{2} & \textbf{3} & \textbf{4} & \textbf{5} &  \\ \hline
\textbf{17 I like to interact with other people in a game.} & 0.907 &  &  &  &  & 0.866 \\ \hline
\textbf{18 I like playing with other people.} & 0.889 &  &  &  &  & 0.834 \\ \hline
\textbf{16 I like to play online with other players.} & 0.887 &  &  &  &  & 0.835 \\ \hline
\textbf{20 I often prefer to play games with others.} & 0.872 &  &  &  &  & 0.794 \\ \hline
\textbf{19 I like games that let me play in guilds or teams.} & 0.858 &  &  &  &  & 0.78 \\ \hline
\textbf{22 I enjoy highly difficult challenges in games.} &  & 0.884 &  &  &  & 0.839 \\ \hline
\textbf{23 I like it when games challenge me.} &  & 0.878 &  &  &  & 0.826 \\ \hline
\textbf{21 I like it when goals are hard to achieve in games.} &  & 0.852 &  &  &  & 0.791 \\ \hline
\textbf{25 I like it when progression in a game demands skill.} &  & 0.83 &  &  &  & 0.727 \\ \hline
\textbf{24 I usually play games at the highest difficulty setting.} &  & 0.768 &  &  &  & 0.624 \\ \hline
\textbf{7 Story is important to me when I play games.} &  &  & 0.885 &  &  & 0.825 \\ \hline
\textbf{9 I like games that pull me in with their story.} &  &  & 0.835 &  &  & 0.76 \\ \hline
\textbf{8 I enjoy complex narratives in a game.} &  &  & 0.82 &  &  & 0.755 \\ \hline
\textbf{10 I feel like storytelling does not get in the way of actually playing the game.} &  &  & 0.745 &  &  & 0.569 \\ \hline
\textbf{6 I usually don't skip the story portions or the cutscenes when I am playing.} &  &  & 0.724 &  &  & 0.586 \\ \hline
\textbf{13 I like completing games 100\%.} &  &  &  & 0.865 &  & 0.767 \\ \hline
\textbf{11 I like to complete all the tasks and objectives in a game.} &  &  &  & 0.814 &  & 0.672 \\ \hline
\textbf{15 I like finishing quests.} &  &  &  & 0.668 &  & 0.507 \\ \hline
\textbf{14 I feel stressed if I do not complete all the tasks in a game.} &  &  &  & 0.665 &  & 0.462 \\ \hline
\textbf{12 I usually do care if I do not complete all optional parts of a game.} &  &  &  & 0.612 &  & 0.39 \\ \hline
\textbf{3 I like to spend some time exploring the game world.} &  &  &  &  & 0.743 & 0.622 \\ \hline
\textbf{5 I often feel in awe with the landscapes or other game imagery.} &  &  &  &  & 0.723 & 0.561 \\ \hline
\textbf{2 I like games with detailed worlds or universes to explore.} &  &  &  &  & 0.717 & 0.61 \\ \hline
\textbf{1 I like games which make me feel like I am actually in a different place.} &  &  &  &  & 0.668 & 0.46 \\ \hline
\textbf{4 I like to customize how my character looks in a game.} &  &  &  &  & 0.601 & 0.391 \\ \hline
\multicolumn{1}{|r|}{\textbf{Variance explained cumulatively \%}} & 16.596 & 32.066 & 45.836 & 58.876 & 67.416 &  \\ \hline
\end{tabular}%
}
\caption{EFA Results of Traits Section in \pogo{}}
\label{tab:pokego-efa}
\end{table}

\paragraph{Results of CFA}

According to the aforementioned criteria regarding confirmatory factor analysis, the following results demonstrated that the factorial validity of the Traits section in \pogo{} survey are acceptable. 
To be specific, the chi square to df ratio ($\chi$2/df) is less than 5, the Tucker Lewis index (TLI) and the Comparative Fit Index (CFI) are all higher than 0.90, the Root Mean Square Error of Approximation (RMSEA) is less than 0.1 and the Standardized Root Mean Square Residual (SRMR) is less than 0.08

\begin{figure}[htpb]
    \centering
    \includegraphics[scale=.60]{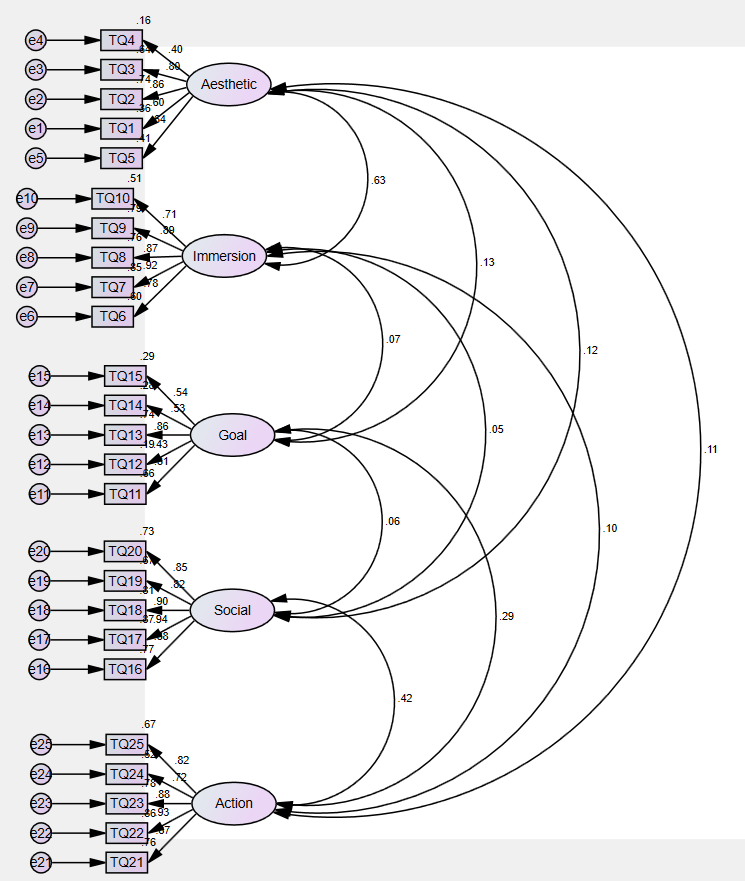}
    \caption{\pogo{} CFA Traits Model with Results}
    \label{fig:pokego-cfa}
\end{figure}

\renewcommand{\thefootnote}{\fnsymbol{footnote}}
\begin{table}[htpb]
\begin{tabular}{|l|l|l|l|l|l|l|}
\hline
\textbf{$\chi$2} & \textbf{df} & \textbf{$\chi$2/df} & \textbf{SRMR} & \textbf{RMSEA} & \textbf{TLI} & \textbf{CFI} \\ \hline
715.603 & 265 & 2.700\footnote[7]{indicates an acceptable fit} & .0513\footnotemark[7] & .056\footnotemark[7] & .942\footnotemark[7] & 0.949\footnotemark[7] \\ \hline
\end{tabular}
\caption{\pogo{} CFA Traits Results}
\label{tab:pokegotraits-cfa}
\end{table}

\paragraph{Results of Reliability}
Based on the aforementioned criteria regarding reliability, our results indicated that the dimensions in the Traits section were reliable as their Cronbach’s alpha were all over the acceptable threshold (0.6) with range from 0.758 to 0.944. \pagebreak

\begin{table}[htpb]
\begin{tabular}{|l|l|l|l|l|l|}
\hline
\textbf{\textbf{Dimension}} & \textbf{\textbf{Aesthetic Ori}} & \textbf{\textbf{Immersion Ori}} & \textbf{\textbf{Goal Ori}} & \textbf{\textbf{Social Ori}} & \textbf{\textbf{Action Ori}} \\ \hline
Cronbach’s $\alpha$ & .758 & .897 & .758 & .944 & .914 \\ \hline
\end{tabular}
\caption{\pogo{} Results of Reliability}
\label{tab:pokego-traits-reliability}
\end{table}

\subsubsection{\hp{}}
\paragraph{Results of the EFA}

Similar to \pogo{}, KMO and Bartlett’s test indicate that the data in Traits section in \hp{} Survey is suitable for exploratory factor analysis.

\begin{longtable}[htpb]{|l|l|l|}
\hline
\multicolumn{3}{|c|}{\textbf{KMO and Bartlett's Test}} \\ \hline
\endfirsthead
\endhead
\multicolumn{2}{|l|}{Kaiser-Meyer-Olkin Measure of Sampling Adequacy} & \textbf{.878} \\ \hline
\multirow{3}{*}{Bartlett's Test of Sphericity} & Approx. Chi-Square & 9612.987 \\ \cline{2-3} 
 & df & 300 \\ \cline{2-3} 
 & Sig. & \textbf{.000} \\ \hline
\caption{Harry Potter KMO and Bartlett's Test Results}
\label{tab:hp-traits-kmobartletts}
\end{longtable}

The Rotated Component Matrix table for\hp{} (Table \ref{tab:hp-efa}) indicates that no problematic item should be removed based on the aforementioned criteria and, consequently, all the 25 items proposed in original Traits questionnaire were remained in our survey. 
In addition, the extracted factors as well as the items involved in each factor were perfectly in line with the findings from Tondello \cite{tondello2019dynamic}. 
Therefore, the five factors are: Aesthetic orientation (Q1-Q5), Immersion orientation (Q6-Q10), Goal orientation (Q11-Q15), Social orientation (Q16-Q20) and Action orientation (Q21-Q25). The five factors explained 71.481\% of the variance cumulatively.

\begin{table}[htpb]
\resizebox{\textwidth}{!}{%
\begin{tabular}{|p{6.5cm}|l|l|l|l|l|l|}
\hline
\multicolumn{1}{|c|}{\multirow{2}{*}{\textbf{Items}}} & \multicolumn{5}{c|}{\textbf{Component}} & \multirow{2}{*}{\textbf{Communalities}} \\ \cline{2-6}
\multicolumn{1}{|c|}{} & \textbf{1} & \textbf{2} & \textbf{3} & \textbf{4} & \textbf{5} &  \\ \hline
\textbf{18 I like playing with other people.} & .921 &  &  &  &  & .878 \\ \hline
\textbf{17 I like to interact with other people in a game.} & .916 &  &  &  &  & .886 \\ \hline
\textbf{16 I like to play online with other players.} & .885 &  &  &  &  & .838 \\ \hline
\textbf{20 I often prefer to play games with others.} & .878 &  &  &  &  & .817 \\ \hline
\textbf{19 I like games that let me play in guilds or teams.} & .873 &  &  &  &  & .824 \\ \hline
\textbf{7 Story is important to me when I play games.} &  & .909 &  &  &  & .884 \\ \hline
\textbf{8 I enjoy complex narratives in a game.} &  & .897 &  &  &  & .861 \\ \hline
\textbf{9 I like games that pull me in with their story.} &  & .862 &  &  &  & .815 \\ \hline
\textbf{6 I usually dont skip the story portions or the cutscenes when I am playing.} &  & .841 &  &  &  & .716 \\ \hline
\textbf{10 I feel like storytelling does not get in the way of actually playing the game.} &  & .789 &  &  &  & .650 \\ \hline
\textbf{22 I enjoy highly difficult challenges in games.} &  &  & .894 &  &  & .846 \\ \hline
\textbf{21 I like it when goals are hard to achieve in games} &  &  & .859 &  &  & .782 \\ \hline
\textbf{23 I like it when games challenge me.} &  &  & .851 &  &  & .784 \\ \hline
\textbf{24 I usually play games at the highest difficulty setting} &  &  & .813 &  &  & .701 \\ \hline
\textbf{25 I like it when progression in a game demands skill.} &  &  & .794 &  &  & .665 \\ \hline
\textbf{13 I like completing games 100\%.} &  &  &  & .870 &  & .789 \\ \hline
\textbf{11 I like to complete all the tasks and objectives in a game.} &  &  &  & .842 &  & .728 \\ \hline
\textbf{14 I feel stressed if I do not complete all the tasks in a game.} &  &  &  & .765 &  & .594 \\ \hline
\textbf{15 I like finishing quests.} &  &  &  & .734 &  & .590 \\ \hline
\textbf{12 I usually do care if I do not complete all optional parts of a game.} &  &  &  & .577 &  & .352 \\ \hline
\textbf{1 I like games which make me feel like I am actually in a different place.} &  &  &  &  & .769 & .602 \\ \hline
\textbf{3 I like to spend some time exploring the game world.} &  &  &  &  & .763 & .681 \\ \hline
\textbf{2 I like games with detailed worlds or universes to explore.} &  &  &  &  & .758 & .675 \\ \hline
\textbf{5 I often feel in awe with the landscapes or other game imagery.} &  &  &  &  & .691 & .524 \\ \hline
\textbf{4 I like to customize how my character looks in a game.} &  &  &  &  & .524 & .388 \\ \hline
\multicolumn{1}{|r|}{\textbf{Variance explained cumulatively \%}} & 16.917 & 32.612 & 48.275 & 60.368 & 71.481 &  \\ \hline
\end{tabular}%
}
\caption{EFA Results of Traits Section in \hp{}}
\label{tab:hp-efa}
\end{table}

\paragraph{Results of the CFA}

\begin{figure}[htpb]
    \centering
    \includegraphics[scale=.60]{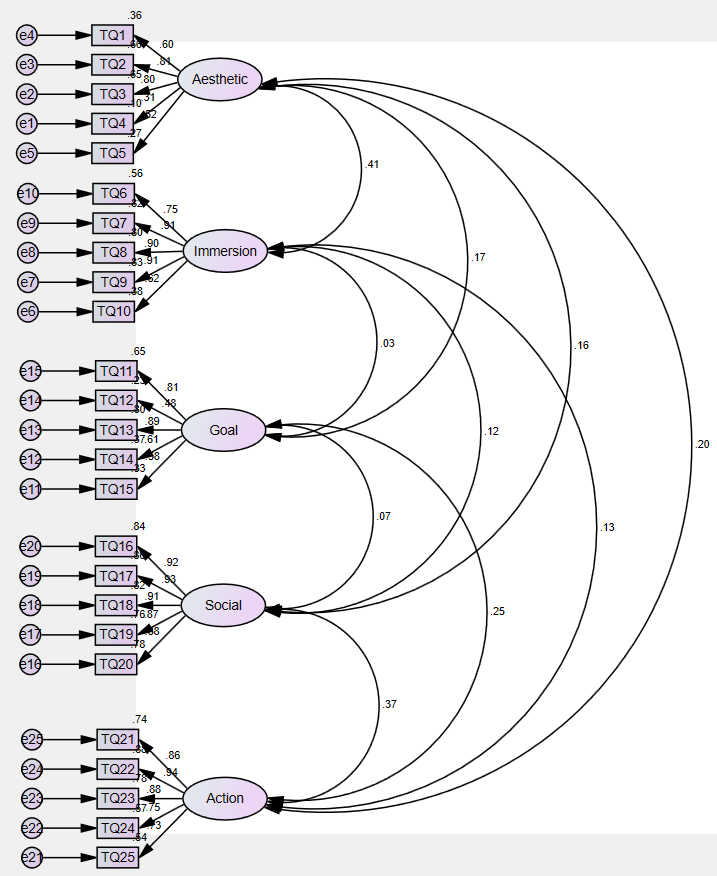}
    \caption{Harry Potter CFA Traits with Results}
    \label{fig:hptraits-cfa}
\end{figure}

Besides the results of EFA which indicated a good construct validity, we further confirmed that the factorial validity of Traits in \hp{} was ideal.
This is reflected in Table \ref{fig:hptraits-cfa} which indicates that all the index reach a good fit of the model to the data. 

\renewcommand{\thefootnote}{\fnsymbol{footnote}}
\begin{table}[htpb]
\begin{tabular}{|l|l|l|l|l|l|l|}
\hline
\textbf{$\chi$2} & \textbf{df} & \textbf{$\chi$2/df} & \textbf{SRMR} & \textbf{RMSEA} & \textbf{TLI} & \textbf{CFI} \\ \hline
711.638 & 265 & 2.685\footnote[7]{indicates an acceptable fit} & .0537\footnotemark[7] & .056\footnotemark[7] & .943\footnotemark[7] & 0.949\footnotemark[7] \\ \hline
\end{tabular}
\caption{\hp{} CFA Traits Results}
\label{tab:hptraits-cfa}
\end{table}

\paragraph{Results of Reliability}

The Cronbach’s alphas displayed in Table \ref{tab:hp-traits-reliability} suggest that all of the dimensions (factors) in Traits section in \hp{} were reliable given the lowest value was beyond the acceptable threshold (0.6) at 0.735 while 3 out of 5 dimensions had excellent reliabilities with Cronbach’s alphas more than 0.9.

\begin{table}[htpb]
\begin{tabular}{|l|l|l|l|l|l|}
\hline
\textbf{\textbf{Dimension}} & \textbf{\textbf{Aesthetic Ori}} & \textbf{\textbf{Immersion Ori}} & \textbf{\textbf{Goal Ori}} & \textbf{\textbf{Social Ori}} & \textbf{\textbf{Action Ori}} \\ \hline
Cronbach’ $\alpha$ & .758 & .897 & .758 & .944 & .914 \\ \hline
\end{tabular}
\caption{\hp{} Results of Reliability}
\label{tab:hp-traits-reliability}
\end{table}

\subsubsection{\ingress{}}
\paragraph{Results of the EFA}

Based on the KMO and Bartlett’s Test, the Traits of \ingress{} survey was also suitable for explorative factor analysis as KMO was larger than 0.6 and Bartlett’s Test of Sphericity was statistically significant with p value equals to .000.
These results can be seen in Table \ref{tab-ingress-kmo}.

\begin{longtable}[t]{|l|l|l|}
\hline
\multicolumn{3}{|c|}{\textbf{KMO and Bartlett's Test}} \\ \hline
\endfirsthead
\endhead
\multicolumn{2}{|l|}{Kaiser-Meyer-Olkin Measure of Sampling Adequacy} & \textbf{.843} \\ \hline
\multirow{3}{*}{Bartlett's Test of Sphericity} & Approx. Chi-Square & 2605.331 \\ \cline{2-3} 
 & df & 276 \\ \cline{2-3} 
 & Sig. & \textbf{.000} \\ \hline
\caption{\ingress{} KMO and Bartlett's Test Results}
\label{tab-ingress-kmo}
\end{longtable}

In the exploratory factor analysis of Traits section in \ ingress{} survey, Q5 - “I often feel in awe with the landscapes or other game imagery” -  should be removed during the test as it was found to load on two factors, which is not permitted based on the aforementioned criteria regarding item retention (see section \ref{CFA-EFA}). 
The results of the EFA presented in Table \ref{tab:ingress-traits-cfa-questions} suggest that the rest items remain a stable 5-factor construct, in which all the items load on expected factors as Tondello \cite{tondello2019dynamic} suggested.
Therefore, for the sake of consistency in this report, the five factors are labeled as Aesthetic orientation (Q1-Q4, Q5 removed), Immersion orientation (Q6-Q10), Goal orientation (Q11-Q15), Social orientation (Q16-Q20) and Action orientation (Q21-Q25). 
The five factors explained 75.698\% of the variances.

\begin{table}[htpb]
\resizebox{\textwidth}{!}{%
\begin{tabular}{|p{6.5cm}|l|l|l|l|l|l|}
\hline
\multicolumn{1}{|c|}{\multirow{2}{*}{\textbf{Items}}} & \multicolumn{5}{c|}{\textbf{Component}} & \multirow{2}{*}{\textbf{Communalities}} \\ \cline{2-6}
\multicolumn{1}{|c|}{} & \multicolumn{1}{c|}{\textbf{1}} & \multicolumn{1}{c|}{\textbf{2}} & \multicolumn{1}{c|}{\textbf{3}} & \multicolumn{1}{c|}{\textbf{4}} & \multicolumn{1}{c|}{\textbf{5}} &  \\ \hline
\textbf{18 I like playing with other people.} & .941 &  &  &  &  & .912 \\ \hline
\textbf{17 I like to interact with other people in a game.} & .935 &  &  &  &  & .900 \\ \hline
\textbf{20 I often prefer to play games with others.} & .896 &  &  &  &  & .855 \\ \hline
\textbf{19 I like games that let me play in guilds or teams.} & .888 &  &  &  &  & .839 \\ \hline
\textbf{16 I like to play online with other players.} & .833 &  &  &  &  & .810 \\ \hline
\textbf{22 I enjoy highly difficult challenges in games.} &  & .888 &  &  &  & .864 \\ \hline
\textbf{23 I like it when games challenge me.} &  & .859 &  &  &  & .797 \\ \hline
\textbf{21 I like it when goals are hard to achieve in games.} &  & .852 &  &  &  & .802 \\ \hline
\textbf{25 I like it when progression in a game demands skill.} &  & .830 &  &  &  & .726 \\ \hline
\textbf{24 I usually play games at the highest difficulty setting.} &  & .820 &  &  &  & .715 \\ \hline
\textbf{6 I usually dont skip the story portions or the cutscenes when I am playing.} &  &  & .880 &  &  & .816 \\ \hline
\textbf{7 Story is important to me when I play games.} &  &  & .863 &  &  & .843 \\ \hline
\textbf{8 I enjoy complex narratives in a game.} &  &  & .855 &  &  & .832 \\ \hline
\textbf{9 I like games that pull me in with their story.} &  &  & .814 &  &  & .863 \\ \hline
\textbf{10 I feel like storytelling does not get in the way of actually playing the game.} &  &  & .750 &  &  & .626 \\ \hline
\textbf{13 I like completing games 100\%.} &  &  &  & .844 &  & .812 \\ \hline
\textbf{11 I like to complete all the tasks and objectives in a game.} &  &  &  & .831 &  & .730 \\ \hline
\textbf{14 I feel stressed if I do not complete all the tasks in a game.} &  &  &  & .735 &  & .621 \\ \hline
\textbf{15 I like finishing quests.} &  &  &  & .680 &  & .612 \\ \hline
\textbf{12 I usually do care if I do not complete all optional parts of a game.} &  &  &  & .595 &  & .400 \\ \hline
\textbf{2 I like games with detailed worlds or universes to explore.} &  &  &  &  & .856 & .792 \\ \hline
\textbf{3 I like to spend some time exploring the game world.} &  &  &  &  & .841 & .769 \\ \hline
\textbf{1 I like games which make me feel like I am actually in a different place.} &  &  &  &  & .815 & .767 \\ \hline
\textbf{4 I like to customize how my character looks in a game.} &  &  &  &  & .607 & .468 \\ \hline
\multicolumn{1}{|r|}{\textbf{Variance explained cumulatively \%}} & 18.394 & 34.814 & 50.938 & 63.373 & 75.698 &  \\ \hline
\end{tabular}%
}
\caption{EFA Results of Traits Section in \ingress{}}
\label{tab:ingress-traits-cfa-questions}
\end{table}

\paragraph{Results of the CFA}

\begin{figure}[htpb]
    \centering
    \includegraphics[scale=.60]{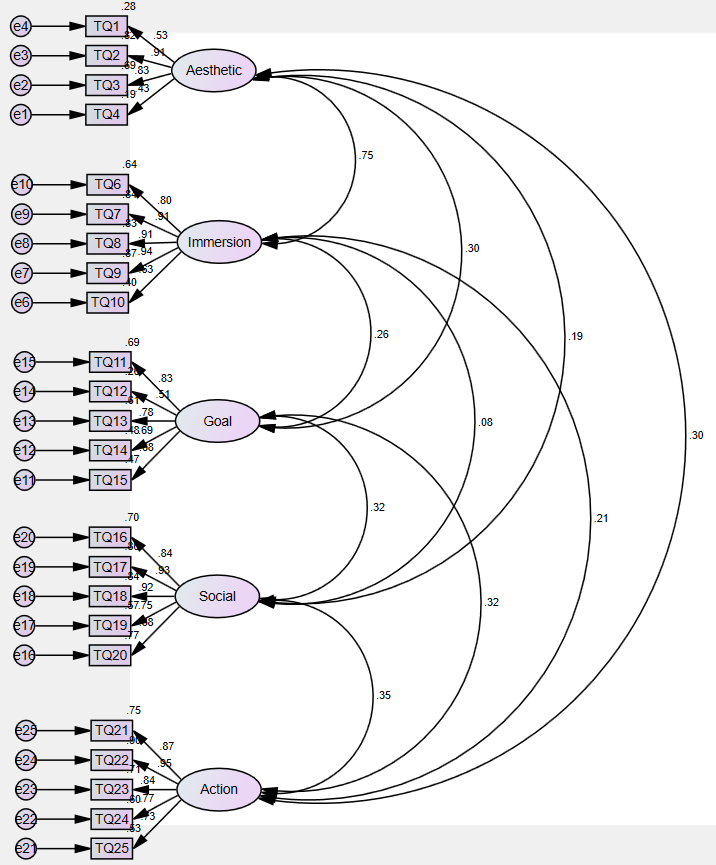}
    \caption{\ingress{} CFA Traits Model with Results}
    \label{fig:ingress-traits-cfa}
\end{figure}

Although the removed Q5 could potentially confound the factorial validity of the Traits section in \ingress{} survey, we confirmed that the questionnaire without Q5 is still valid for further analysis.
The following statistics reveal that $\chi$2/df, SRMR, TLI and CFI are all good and RSMEA was also within the acceptable threshold.
These results can be see in Figure \ref{fig:ingress-traits-cfa} and Table \ref{tab:ingress-traits-cfa-overall}.

\begin{table}[htpb]
\begin{tabular}{|l|l|l|l|l|l|l|}
\hline
\textbf{$\chi$2} & \textbf{df} & \textbf{$\chi$2/df} & \textbf{SRMR} & \textbf{RMSEA} & \textbf{TLI} & \textbf{CFI} \\ \hline
451.589 & 242 & 1.866\footnote[7]{indicates an acceptable fit} & .0732\footnotemark[7] & .082\footnotemark[7] & .896\footnotemark[7] & 0.909\footnotemark[7] \\ \hline
\end{tabular}
\caption{\ingress{} CFA Traits Results}
\label{tab:ingress-traits-cfa-overall}
\end{table}

\paragraph{Results of Reliability}
The results of Cronbach’s alpha also indicate that all the dimensions (factors) in the Traits section in \ingress{} survey were reliable ranging from 0.780 to 0.944.
These results can be seen in Table \ref{tab:ingress-traits-reliability}

\begin{table}[htpb]
\begin{tabular}{|l|l|l|l|l|l|}
\hline
\textbf{\textbf{Dimension}} & \textbf{\textbf{Aesthetic Ori}} & \textbf{\textbf{Immersion Ori}} & \textbf{\textbf{Goal Ori}} & \textbf{\textbf{Social Ori}} & \textbf{\textbf{Action Ori}} \\ \hline
Cronbach’ $\alpha$ & .758 & .897 & .758 & .944 & .914 \\ \hline
\end{tabular}
\caption{\ingress{} Results of Reliability}
\label{tab:ingress-traits-reliability}
\end{table}

\subsection{Reliability and Validity of Gratifications Section of Surveys}

In order to explore the gratifications that the casual and hardcore players derive from playing \pogo{}, \hp{}, and \ingress{}, we utilized 33 items in 7 dimensions from Hamari’s scale  \cite{hamari2019uses}, namely Challenge, Competition, Enjoyment, Trendiness, Socializing, Outdoor activity and Nostalgia.

\subsubsection{\pogo{}}
\paragraph{Results of the EFA}

The results of KMO and Bartlett’s test (Table \ref{tab:pokego-efa-gratification}) indicate that the data in Gratification section in \pogo{} is suitable for exploratory factor analysis. 
To be specific, the KMO equals to 0.890 suggest that the sample is large enough to perform EFA. 
In addition, the significance (p<0.05) of Bartlett’s of Sphericity indicated that the correlation between items are sufficiently large.

\begin{table}[htpb]
\begin{tabular}{|l|l|l|}
\hline
\multicolumn{2}{|l|}{\textbf{Kaiser-Meyer-Olkin Measure of Sampling Adequacy.}} & .880 \\ \hline
\textbf{Bartlett's Test of Sphericity} & \textbf{Approx. Chi-Square} & 8957.427 \\ \hline
\textbf{} & \textbf{df} & 325 \\ \hline
\textbf{} & \textbf{Sig.} & .000 \\ \hline
\end{tabular}
\caption{\pogo{} KMO and Bartlett's Test Results}
\label{tab:pokego-efa-gratification}
\end{table}

According to the criteria we set to determine the retention of items, we removed 5 out of 33 items in the Gratification section in \pogo{} survey. 
The detailed items as well as the reasons why they were removed are listed in Table \ref{tab:poke-gratifications-questions}:

\begin{table}[htpb]
\begin{tabular}{|p{5cm}|p{5cm}|}
\hline
\textbf{Items} & \textbf{Reasons} \\ \hline
Q14 I play \pogo{} because it is a habit. & A valid factor should have at least 3 items. Q14 is the items with the higher loading in a two-item factor. \\ \hline
Q25 I play \pogo{} because it motivates me to go out. & A valid factor should have at least 3 items. Q25 is the items with the higher loading in a two-item factor. \\ \hline
Q15 I play \pogo{} because it occupies my free time. & Low factor loading and low Communalities \\ \hline
Q3 I feel excited when I catch a new Pok{\'e}mon. & Loaded on 2 or more factors \\ \hline
Q26 I play \pogo{} because it motivates me to explore new places. & Loaded on 2 or more factors \\ \hline
Q6 I like to prove that I am one of the best players. & Loaded on 2 or more factors \\ \hline
Q10 I play \pogo{} because it is exciting. & Loaded on 2 or more factors \\ \hline
\end{tabular}
\caption{Questions Removed from our Survey and reasons for their removal}
\label{tab:poke-gratifications-questions}
\end{table}

The final result of EFA (as seen in Table \ref{tab:pokego-gratification-efa} suggest that a 6-factor structure can be extracted from the data in Gratification section in \pogo{} survey, although we initially employed items from 7 dimensions in the original Gratification questionnaire. 
The difference is that, instead of having Socializing and Outdoor activity dimensions separately, our data suggest that the two dimensions should be merged as one. 
This is reflected in Table \ref{tab:pokego-gratification-efa}. The first extracted factor in our result has 8 items, among which, 6 items (Q19, Q20, Q21, Q22, Q23, Q24) belong to Socializing dimensions and 2 items (Q27, Q28) belong to Outdoor Activity dimensions in original Gratification questionnaire. 
However, this is not surprising as we deployed our questionnaire in new samples who may have different play experiences and the difference in samples would have impact on the structure of factors \cite{gaskin2017sample}. 
In addition, the participants in our sample may interpret questions differently in comparison to those in Hamari \cite{hamari2019uses} because more than half of our participants were native English speakers who came from US and UK while nearly 70\% of the participants in Hamari \cite{hamari2019uses} were non-native English speakers who were mainly from Finland and Philippines. 

More importantly, the Outdoor activity dimension in the original questionnaire includes questions that are closely related to social activities. 
For example, Q27 “I play \pogo{} because I can meet friends outdoors.” and Q28 “I play \pogo{} because I can meet strangers outdoors.”.  
This may have resulted in the two dimensions being interpreted as one by the participants. 
Therefore, after evaluating the EFA results of a 7-factor construct, which showed poor correspondences between items and factors, we decided to accept the 6-factor structure in the Gratification section in \pogo{} survey. 

The first factor is labeled as “Socialization” to distinguish from the “Socializing” dimension in the original Gratification questionnaire.
Socialization describe the player’s gratification-related interactions with others. 
Since the items involved in the rest factors are largely in line with the original questionnaire, we decide to label the remaining factors with the same names that Hamari \cite{hamari2019uses} used in order to keep the consistency in naming. 

In detail, the second factor is named “Enjoyment”, which describes the pleasure an individual gains while playing the game. 
The third factor is labeled as “Trendiness” to indicate the extent to which an individual considers playing game for other’s perceptions. 
The fourth factor describes player’s yearning to relive or return to a past period so that it is named “Nostalgia”. 
Factor 5 is called “Challenge” as it evaluates a sense that an individual’s capabilities are being stretched and tested. 
The last factor describes the desire and need of players to perform better or beat other players/ computer while playing a game so that it is named “Competition”. 
The 6 factors explained 69.872\% of the variance cumulatively (seen in Table \ref{tab:pokego-gratification-efa}.

\begin{table}[htpb]
\resizebox{\textwidth}{!}{%
\begin{tabular}{|p{6.5cm}|l|l|l|l|l|l|l|}
\hline
\multicolumn{1}{|c|}{\multirow{2}{*}{\textbf{Items}}} & \multicolumn{6}{c|}{\textbf{Component}} & \multirow{2}{*}{\textbf{Communalities}} \\ \cline{2-7}
\multicolumn{1}{|c|}{} & \multicolumn{1}{c|}{\textbf{1}} & \multicolumn{1}{c|}{\textbf{2}} & \multicolumn{1}{c|}{\textbf{3}} & \multicolumn{1}{c|}{\textbf{4}} & \multicolumn{1}{c|}{\textbf{5}} & \multicolumn{1}{c|}{\textbf{6}} &  \\ \hline
\textbf{24 \pogo{} enables me to be part of a group.} & .838 &  &  &  &  &  & .761 \\ \hline
\textbf{27 I play \pogo{} because I can meet friends outdoors.} & .807 &  &  &  &  &  & .716 \\ \hline
\textbf{21 \pogo{} enables me to make new friends.} & .798 &  &  &  &  &  & .728 \\ \hline
\textbf{19 \pogo{} enables me to maintain friendships.} & .766 &  &  &  &  &  & .675 \\ \hline
\textbf{20 \pogo{} enables me to improve relationships.} & .763 &  &  &  &  &  & .650 \\ \hline
\textbf{23 \pogo{} enables me to participate in relevant discussions.} & .753 &  &  &  &  &  & .655 \\ \hline
\textbf{22 I like to play \pogo{} because my friends play the game.} & .707 &  &  &  &  &  & .546 \\ \hline
\textbf{28 I play \pogo{} because I can meet strangers outdoors.} & .683 &  &  &  &  &  & .617 \\ \hline
\textbf{11 I play \pogo{} because it is entertaining.} &  & .888 &  &  &  &  & .851 \\ \hline
\textbf{12 I play \pogo{} because it is fun.} &  & .869 &  &  &  &  & .836 \\ \hline
\textbf{13 I play \pogo{} because it is a good pastime.} &  & .753 &  &  &  &  & .632 \\ \hline
\textbf{10 I play \pogo{} because it is exciting.} &  & .660 &  &  &  &  & .649 \\ \hline
\textbf{18 \pogo{} enables me to look stylish.} &  &  & .922 &  &  &  & .958 \\ \hline
\textbf{16 \pogo{} enables me to look trendy.} &  &  & .912 &  &  &  & .920 \\ \hline
\textbf{17 \pogo{} enables me to look cool.} &  &  & .912 &  &  &  & .945 \\ \hline
\textbf{33 I have been a fan of Pok{\'e}mon even before the launch of \pogo{}.} &  &  &  & .867 &  &  & .757 \\ \hline
\textbf{31 I used to watch Pok{\'e}mon cartoons/anime series/movies.} &  &  &  & .805 &  &  & .652 \\ \hline
\textbf{30 I used to play Pok{\'e}mon games on Nintendo Gameboy or other consoles.} &  &  &  & .785 &  &  & .622 \\ \hline
\textbf{32 I used to collect Pok{\'e}mon merchandise (e.g. toys, stickers, trading cards, books etc.).} &  &  &  & .773 &  &  & .610 \\ \hline
\textbf{4 I feel excited when I win a battle.} &  &  &  &  & .709 &  & .593 \\ \hline
\textbf{5 I enjoy finding new and creative ways to work through \pogo{}.} &  &  &  &  & .673 &  & .585 \\ \hline
\textbf{1 I feel proud when I master an aspect of \pogo{}.} &  &  &  &  & .634 &  & .576 \\ \hline
\textbf{2 It feels rewarding to get to the next level.} &  &  &  &  & .610 &  & .538 \\ \hline
\textbf{8 I get upset when I am unable to earn enough points.} &  &  &  &  &  & .813 & .718 \\ \hline
\textbf{7 I get upset when others do better than me.} &  &  &  &  &  & .810 & .725 \\ \hline
\textbf{9 It is important to me to be one of the skilled persons playing the game.} &  &  &  &  &  & .708 & .653 \\ \hline
\multicolumn{1}{|r|}{\textbf{Variance explained cumulatively \%}} & 19.843 & 31.379 & 42.071 & 52.360 & 61.634 & 69.872 &  \\ \hline
\end{tabular}%
}
\caption{EFA Results of Gratification Section in \pogo{}}
\label{tab:pokego-gratification-efa}
\end{table}

\paragraph{Results of the CFA}

\begin{figure}[htpb]
    \centering
    \includegraphics[scale=.60]{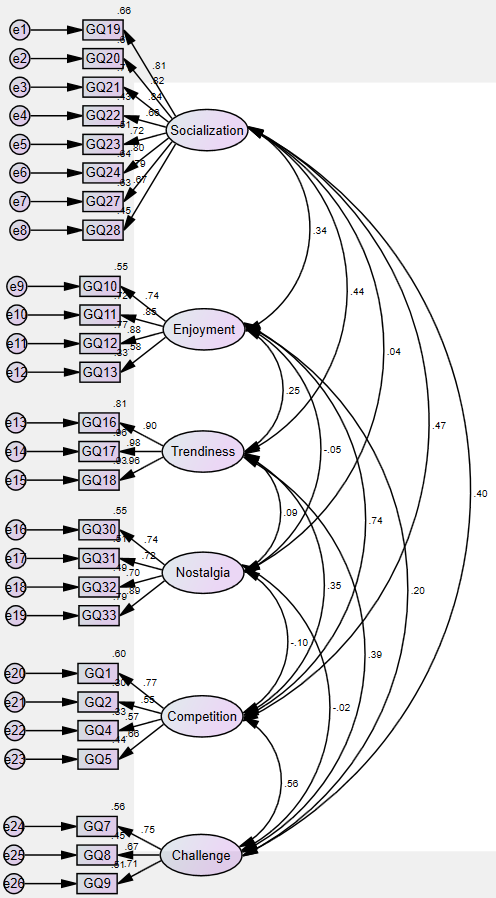}
    \caption{\pogo{} CFA Gratifications with Results}
    \label{fig:poke-gratifications-cfa}
\end{figure}

According to the aforementioned criteria regarding confirmatory factor analysis, the results (located in Figure \ref{fig:poke-gratifications-cfa} and Table \ref{tab:poke-gratifications-cfa}) demonstrate that the factorial validity of the Gratification section in \pogo{} survey was acceptable.
This acceptability is due to the $\chi$2/df value being acceptable at 3.552, both the Tucker Lewis index (TLI) and the Comparative Fit Index (CFI) resulting in values over 0.9, and both the Root Mean Square Error of Approximation (RMSEA) and the Standardized Root Mean Square Residual (SRMR) resulting in values less than 0.08.

\begin{table}[htpb]
\begin{tabular}{|l|l|l|l|l|l|l|}
\hline
\textbf{$\chi$2} & \textbf{df} & \textbf{$\chi$2/df} & \textbf{SRMR} & \textbf{RMSEA} & \textbf{TLI} & \textbf{CFI} \\ \hline
1008.888 & 284 & 3.552\footnote[7]{indicates an acceptable fit} & .0600\footnotemark[7] & .069\footnotemark[7] & .903\footnotemark[7] & 0.915\footnotemark[7] \\ \hline
\end{tabular}
\caption{\pogo{} CFA Gratifications Results}
\label{tab:poke-gratifications-cfa}
\end{table}

\paragraph{Results of Reliability}

The statistics in Table \ref{tab:poke-gratifications-reliability} indicate that the Gratification was a reliable instrument as the Cronbach’s alpha in all the six dimensions were more than 0.6 ranging from 0.703 to 0.966.

\begin{table}[htpb]
\resizebox{\textwidth}{!}{%
\begin{tabular}{|l|l|l|l|l|l|l|}
\hline
\textbf{Dimension} & \textbf{Socialization} & \textbf{Enjoyment} & \textbf{Trendiness} & \textbf{Nostalgia} & \textbf{Competition} & \textbf{Challenge} \\ \hline
\textbf{Cronbach’s $\alpha$} & .904 & .849 & .966 & .944 & .703 & .815 \\ \hline
\end{tabular}%
}
\caption{\pogo{} Gratifications Results of Reliability}
\label{tab:poke-gratifications-reliability}
\end{table}

\subsubsection{\hp{}}
\paragraph{Results of the EFA}

The statistics of KMO and Barlett’s test (Table \ref{tab:hp-kmo-efa-gratification}) suggest that data in Gratification section in \hp{} survey is suitable for exploratory factor analysis as KMO was larger than 0.6 and result of Bartlett’s test was significant at p less than 0.05.

\begin{table}[htpb]
\begin{tabular}{|l|l|l|}
\hline
\multicolumn{2}{|l|}{\textbf{Kaiser-Meyer-Olkin Measure of Sampling Adequacy.}} & .882 \\ \hline
\textbf{Bartlett's Test of Sphericity} & \textbf{Approx. Chi-Square} & 10007.259 \\ \hline
\textbf{} & \textbf{df} & 378 \\ \hline
\textbf{} & \textbf{Sig.} & .000 \\ \hline
\end{tabular}
\caption{\hp{} Go KMO and Bartlett's Test Results}
\label{tab:hp-kmo-efa-gratification}
\end{table}

In order to draw a valid factor structure of Gratification section in \hp{} survey, 8 out of 33 items were eliminated during the exploratory factor analysis. 
Details of the removed items are provided in Table \ref{tab:hp-questions-removed}.

\begin{table}[htpb]
\begin{tabular}{|p{4cm}|p{4cm}|}
\hline
\textbf{Items} & \textbf{Reasons} \\ \hline
Q26 I play \hp{} because it motivates me to explore new places. & A valid factor should have at least 3 items. Q26 is the items with the higher loading in a two-item factor. \\ \hline
Q15 I play \hp{} because it occupies my free time. & A valid factor should have at least 3 items. Q15 is the items with the higher loading in a two-item factor. \\ \hline
Q32 I used to collect Harry Potter merchandise (e.g. toys, stickers, trading cards, etc.). & A valid factor should have at least 3 items. Q32 is the items with the higher loading in a two-item factor. \\ \hline
Q25 I play \hp{} because it motivates me to go out. & Low communalities and Low factor loading \\ \hline
Q14 I play \hp{} because it is a habit. & Low factor loading \\ \hline
\end{tabular}
\caption{Questions Removed from the Gratification Survey for Harry Potter Wizard's Unite}
\label{tab:hp-questions-removed}
\end{table}

The final results of EFA, again, suggest that a 6-factor structure in the Gratification section in \hp{} survey.
These results can be seen in Table \ref{tab:hp-efa-gratification}.
This is because that the aforementioned issue in Outdoor activity remains. 
Half of the items in this dimension (Q25, Q26) were removed due to problematic loading. 
The retained items (Q27, Q28) in Outdoor activity dimension together with the 6 items in Socializing dimensions in the original questionnaire (Q19-24) were merged into a new extracted factor that is marked as factor 1 in the following table. 
Therefore, similar to the way we name the factors in \pogo{} survey, we label the first factor as “Socialization” to distinguish it from Socializing dimension and Outdoor activity dimension in the original questionnaire.  
Since the rest of the factors as well as their corresponded items were largely in line with the results in Hamari \cite{hamari2019uses}, factor 2 is named as “Challenge”, factor 3 is named as “Competition”, factor 4 is named as “Enjoyment”, factor 5 is named as “Trendiness” and factor 6 is named as “Nostalgia” (detailed definitions of the factors is provided in section \ref{tg-pogo}). 
The 6 factors explained 68.136\% of the variance in total. 

\begin{table}[htpb]
\resizebox{\textwidth}{!}{%
\begin{tabular}{|p{6.5cm}|l|l|l|l|l|l|l|}
\hline
\multicolumn{1}{|c|}{\multirow{2}{*}{\textbf{Items}}} & \multicolumn{6}{c|}{\textbf{Component}} & \multirow{2}{*}{\textbf{Communalities}} \\ \cline{2-7}
\multicolumn{1}{|c|}{} & \multicolumn{1}{c|}{\textbf{1}} & \multicolumn{1}{c|}{\textbf{2}} & \multicolumn{1}{c|}{\textbf{3}} & \multicolumn{1}{c|}{\textbf{4}} & \multicolumn{1}{c|}{\textbf{5}} & \multicolumn{1}{c|}{\textbf{6}} &  \\ \hline
\textbf{24 \hp{} enables me to be part of a group.} & .803 &  &  &  &  &  & .699 \\ \hline
\textbf{20 \hp{} enables me to improve relationships.} & .796 &  &  &  &  &  & .715 \\ \hline
\textbf{27 I play \hp{} because I can meet friends outdoors} & .791 &  &  &  &  &  & .681 \\ \hline
\textbf{21 \hp{} enables me to make new friends} & .771 &  &  &  &  &  & .655 \\ \hline
\textbf{19 \hp{} enables me to maintain friendships.} & .756 &  &  &  &  &  & .671 \\ \hline
\textbf{22 I like to play \hp{} because my friends play the game.} & .736 &  &  &  &  &  & .551 \\ \hline
\textbf{28 I play \hp{} because I can meet strangers outdoors} & .709 &  &  &  &  &  & .585 \\ \hline
\textbf{23 \hp{} enables me to participate in relevant discussions.} & .695 &  &  &  &  &  & .565 \\ \hline
\textbf{4 I feel excited when I win a battle.} &  & .802 &  &  &  &  & .724 \\ \hline
\textbf{3 I feel excited when I return foundables to their proper location.} &  & .778 &  &  &  &  & .728 \\ \hline
\textbf{2 It feels rewarding to get to the next level.} &  & .681 &  &  &  &  & .618 \\ \hline
\textbf{1 I feel proud when I master an aspect of \hp{}} &  & .566 &  &  &  &  & .619 \\ \hline
\textbf{5 I enjoy finding new and creative ways to work through \hp{}.} &  & .558 &  &  &  &  & .534 \\ \hline
\textbf{9 It is important to me to be one of the skilled persons playing the game.} &  &  & .811 &  &  &  & .749 \\ \hline
\textbf{7 I get upset when others do better than me.} &  &  & .794 &  &  &  & .698 \\ \hline
\textbf{6 I like to prove that I am one of the best players.} &  &  & .783 &  &  &  & .702 \\ \hline
\textbf{8 I get upset when I am unable to earn enough points.} &  &  & .617 &  &  &  & .408 \\ \hline
\textbf{11 I play \hp{} because it is entertaining.} &  &  &  & .834 &  &  & .822 \\ \hline
\textbf{12 I play \hp{} because it is fun.} &  &  &  & .797 &  &  & .823 \\ \hline
\textbf{13 I play \hp{} because it is a good pastime.} &  &  &  & .746 &  &  & .641 \\ \hline
\textbf{10 I play \hp{} because it is exciting.} &  &  &  & .615 &  &  & .672 \\ \hline
\textbf{17 \hp{} enables me to look cool.} &  &  &  &  & .917 &  & .961 \\ \hline
\textbf{18 \hp{} enables me to look stylish.} &  &  &  &  & .902 &  & .945 \\ \hline
\textbf{16 \hp{} enables me to look trendy.} &  &  &  &  & .894 &  & .915 \\ \hline
\textbf{33 I have been a fan of Harry Potter even before the launch of \hp{}.} &  &  &  &  &  & .842 & .725 \\ \hline
\textbf{31 I used to watch Harry Potter movies.} &  &  &  &  &  & .813 & .693 \\ \hline
\textbf{29 I used to read Harry Potter books.} &  &  &  &  &  & .734 & .565 \\ \hline
\multicolumn{1}{|r|}{\textbf{Variance explained cumulatively \%}} & 18.600 & 29.757 & 40.225 & 50.549 & 60.367 & 68.136 &  \\ \hline
\end{tabular}%
}
\caption{EFA Results of Gratification Section in \hp{}}
\label{tab:hp-efa-gratification}
\end{table}

\paragraph{Results of the CFA}

\begin{figure}[htpb]
    \centering
    \includegraphics[scale=.60]{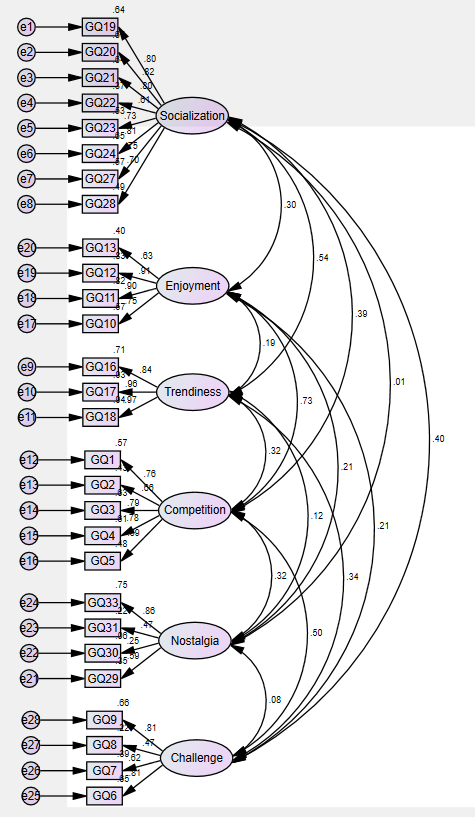}
    \caption{Harry Potter CFA Gratifications with Results}
    \label{fig:hp-cfa-grat}
\end{figure}

After examining the construct validity, the results of CFA confirm that the factorial validity of the Gratification section in Harry Potter Wizard United survey is acceptable.
These results can be seen in Figure \ref{fig:hp-cfa-grat} and Table \ref{tab:hp-cfa-grat}.
The results are acceptable because the Tucker Lewis index (TLI) and the Comparative Fit Index (CFI) are more than the acceptable threshold (0.8). 
The Root Mean Square Error of Approximation (RMSEA) and the Standardized Root Mean Square Residual (SRMR) are clearly within the recommended bounds (0.1 for RMSEA and 0.08 for SRMR).

\begin{table}[htpb]
\begin{tabular}{|l|l|l|l|l|l|l|}
\hline
\textbf{$\chi$2} & \textbf{df} & \textbf{$\chi$2/df} & \textbf{SRMR} & \textbf{RMSEA} & \textbf{TLI} & \textbf{CFI} \\ \hline
1294.885 & 335 & 3.865* & 0.0575** & .074** & .877* & .891* \\ \hline
\end{tabular}
\caption{\hp{} Gratifications CFA Results}
\label{tab:hp-cfa-grat}
\end{table}

\paragraph{Results of Reliability}

Based on the statistics in Table \ref{tab:hp-gratifications-reliability}, it can be safely said that the Gratification section in \hp{} survey is reliable as the Cronbach’s alphas in all the dimensions are beyond the minimum acceptable threshold at 0.6.

\begin{table}[htpb]
\resizebox{\textwidth}{!}{%
\begin{tabular}{|l|l|l|l|l|l|l|}
\hline
\textbf{Dimension} & \textbf{Socialization} & \textbf{Enjoyment} & \textbf{Trendiness} & \textbf{Nostalgia} & \textbf{Competition} & \textbf{Challenge} \\ \hline
\textbf{Cronbach’s $\alpha$} & .904 & .853 & .958 & .944 & .697 & .794 \\ \hline
\end{tabular}%
}
\caption{\hp{} Gratifications Results of Reliability}
\label{tab:hp-gratifications-reliability}
\end{table}

\subsubsection{\ingress{}}
\paragraph{Results of the EFA}

Similar to all the previous sections, the results of KMO and Bartlett’s test in Gratification section in \ingress{} survey indicate that the data was also suitable for exploratory factor analysis, given KMO was 0.882 > 0.6 and result of Bartlett’s test are significant at p<0.05.
These data can be seen in Table \ref{tab:hp-gratifications-kmos}.

\begin{longtable}[htpb]{|l|l|l|}
\hline
\multicolumn{3}{|c|}{\textbf{KMO and Bartlett's Test}} \\ \hline
\endfirsthead
\endhead
\multicolumn{2}{|l|}{Kaiser-Meyer-Olkin Measure of Sampling Adequacy} & \textbf{.853} \\ \hline
\multirow{3}{*}{Bartlett's Test of Sphericity} & Approx. Chi-Square & 2083.636 \\ \cline{2-3} 
 & df & 210 \\ \cline{2-3} 
 & Sig. & \textbf{.000} \\ \hline
\caption{\ingress{} Gratifications Bartlett's Test Results}
\label{tab:hp-gratifications-kmos}
\end{longtable}

Notably, since there was no \ingress{}-related merchandise, cartoon or console games on the market prior to the release of \ingress{}, we excluded the 5 items in Nostalgia dimensions in this test. 
Therefore, in total 28 items from 6 dimensions in the original questionnaire were involved in the exploratory factor analysis conducted on the Gratification section in \ingress{} survey. 
The 6 original dimensions are: Challenge, Competition, Enjoyment, Trendiness, Socializing and Outdoor activity. 
In order to examine the valid construct in this section, 6 out of 28 items were removed. 
The removed items can be seen in Table \ref{tab:ingress-gratifications-questions}.

\begin{table}[htpb]
\begin{tabular}{|p{6.5cm}|p{6.5cm}|}
\hline
\textbf{Items} & \textbf{Reasons} \\ \hline
Q15 I play \ingress{} because it occupies my free time. & A valid factor should have at least 3 items. Q15 is the items with the higher loading in a two-item factor. \\ \hline
Q26 I play \ingress{} because it motivates me to explore new places. & A valid factor should have at least 3 items. Q26 is the items with the higher loading in a two-item factor. \\ \hline
Q14 I play \ingress{} because it is a habit. & Low factor loading \\ \hline
Q28 I play \ingress{} because I can meet strangers outdoors. & Low factor loading \\ \hline
Q10 I play \ingress{} because it is exciting. & Loaded on 2 or more factors \\ \hline
Q25 I feel excited when I return foundables to their proper location. & Loaded on 2 or more factors \\ \hline
Q6 I like to prove that I am one of the best players. & Loaded on 2 or more factors \\ \hline
\end{tabular}
\caption{Questions removed from the \ingress{} Gratifications Survey}
\label{tab:ingress-gratifications-questions}
\end{table}
\pagebreak

Unsurprisingly, the items in Socializing dimension and Outdoor activity are found to be merged again while the rest extracted factors were similar to the results in Hamari \cite{hamari2019uses}.
Please view the results in Table \ref{tab:ingress-efa-grat}.
Therefore, a five-factor structure was extracted from the data in the Gratification section in \ingress{} survey. 
The extracted factors are: 1.) Socialization; 2.) Challenge; 3.) Trendiness; 4.) Enjoyment and 5.) Competition. 
The detailed definition of the 5 factors are provided in section \ref{tg-pogo}. 71.730\% of the variance is explained by the factors cumulatively.

\begin{table}[htbp]
\resizebox{\textwidth}{!}{%
\begin{tabular}{|p{6.5cm}|l|l|l|l|l|l|}
\hline
\multicolumn{1}{|c|}{\multirow{2}{*}{\textbf{Items}}} & \multicolumn{5}{c|}{\textbf{Component}} & \multirow{2}{*}{\textbf{Communalities}} \\ \cline{2-6}
\multicolumn{1}{|c|}{} & \multicolumn{1}{c|}{\textbf{1}} & \multicolumn{1}{c|}{\textbf{2}} & \multicolumn{1}{c|}{\textbf{3}} & \multicolumn{1}{c|}{\textbf{4}} & \multicolumn{1}{c|}{\textbf{5}} &  \\ \hline
\textbf{19 \ingress{} enables me to maintain friendships.} & .866 &  &  &  &  & .780 \\ \hline
\textbf{20 \ingress{} enables me to improve relationships.} & .838 &  &  &  &  & .748 \\ \hline
\textbf{23 \ingress{} enables me to participate in relevant discussions.} & .764 &  &  &  &  & .682 \\ \hline
\textbf{21 \ingress{} enables me to make new friends.} & .748 &  &  &  &  & .738 \\ \hline
\textbf{24 \ingress{} enables me to be part of a group.} & .722 &  &  &  &  & .671 \\ \hline
\textbf{27 I play \ingress{} because I can meet friends outdoors.} & .702 &  &  &  &  & .615 \\ \hline
\textbf{22 I like to play \ingress{} because my friends play the game.} & .677 &  &  &  &  & .525 \\ \hline
\textbf{25 I play \ingress{} because it motivates me to go out.} & .450 &  &  &  &  & .467 \\ \hline
\textbf{2 It feels rewarding to get to the next level.} &  & .845 &  &  &  & .774 \\ \hline
\textbf{1 I feel proud when I master an aspect of \ingress{}.} &  & .787 &  &  &  & .730 \\ \hline
\textbf{5 I enjoy finding new and creative ways to work through \ingress{}.} &  & .688 &  &  &  & .650 \\ \hline
\textbf{4 I feel excited when I win a battle.} &  & .629 &  &  &  & .543 \\ \hline
\textbf{17 \ingress{} enables me to look cool.} &  &  & .944 &  &  & .951 \\ \hline
\textbf{18 \ingress{} enables me to look stylish.} &  &  & .941 &  &  & .947 \\ \hline
\textbf{16 \ingress{} enables me to look trendy.} &  &  & .851 &  &  & .803 \\ \hline
\textbf{11 I play \ingress{} because it is entertaining.} &  &  &  & .840 &  & .811 \\ \hline
\textbf{13 I play \ingress{} because it is a good pastime.} &  &  &  & .804 &  & .786 \\ \hline
\textbf{12 I play \ingress{} because it is fun.} &  &  &  & .804 &  & .825 \\ \hline
\textbf{8 I get upset when I am unable to earn enough points.} &  &  &  &  & .861 & .769 \\ \hline
\textbf{7 I get upset when others do better than me.} &  &  &  &  & .835 & .782 \\ \hline
\textbf{9 It is important to me to be one of the skilled persons playing the game.} &  &  &  &  & .540 & .467 \\ \hline
\multicolumn{1}{|r|}{\textbf{Variance explained cumulatively \%}} & 22.371 & 36.111 & .49.587 & 62.267 & 71.730 &  \\ \hline
\end{tabular}%
}
\caption{EFA Results of Gratification Section in \ingress{}}
\label{tab:ingress-efa-grat}
\end{table}

\paragraph{Results of the CFA}

\begin{figure}[htpb]
    \centering
    \includegraphics[scale=.60]{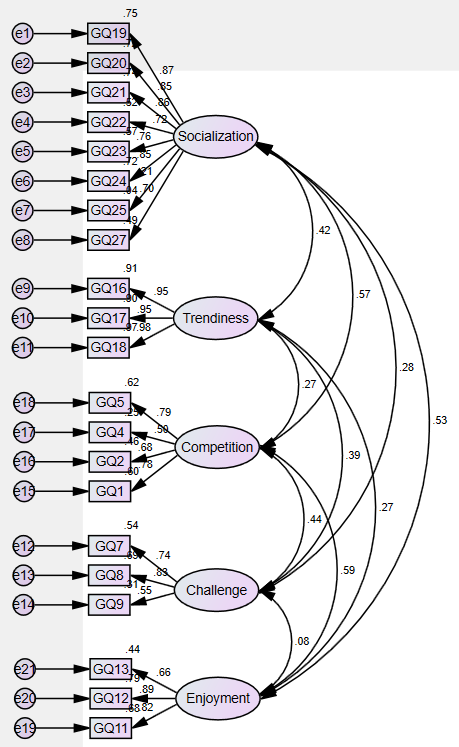}
    \caption{\ingress{} CFA Gratifications with Results}
    \label{fig:ingress-cfa-grats}
\end{figure}

The results of confirmatory factor analysis displayed in Figure \ref{fig:ingress-cfa-grats} and Table \ref{tab:Ingress-cfa-grats} reveal the chi square to degree of freedom ratio ($\chi$2/df), Standardized Root Mean Square Residual (SRMR), Root Mean Square Error of Approximation (RMSEA), Goodness-of-Fit Index (TLI) and Comparative fit index (CFI) were all acceptable according to the aforementioned criteria (see section \ref{CFA-EFA}). 
Therefore, the factorial validity of the Gratification section in \ingress{} survey is confirmed.

\renewcommand{\thefootnote}{\fnsymbol{footnote}}
\begin{table}[htpb]
\begin{tabular}{|l|l|l|l|l|l|l|}
\hline
\textbf{$\chi$2} & \textbf{df} & \textbf{$\chi$2/df} & \textbf{SRMR} & \textbf{RMSEA} & \textbf{TLI} & \textbf{CFI} \\ \hline
348.209 & 179 & 1.945\footnote[7]{indicates an acceptable fit} & .0871\footnotemark[7] & .085\footnotemark[7] & .894\footnotemark[7] & 0.909\footnotemark[7] \\ \hline
\end{tabular}
\caption{\ingress{} CFA Gratifications Results}
\label{tab:Ingress-cfa-grats}
\end{table}

\paragraph{Results of Reliability}

The statistics in Table \ref{tab:Ingress-reliability-grats} indicate that the Gratification questionnaire are reliable as the Cronbach’s alphas in all the dimensions are beyond the acceptable threshold ($\alpha$> 0.6) ranging from 0.724 to 0.962.

\begin{table}[htpb]
\begin{tabular}{|l|l|l|l|l|l|}
\hline
\textbf{Dimension} & \textbf{Socialization} & \textbf{Enjoyment} & \textbf{Trendiness} & \textbf{Competition} & \textbf{Challenge} \\ \hline
Cronbach’s $\alpha$ & .922 & .858 & .962 & .724 & .733 \\ \hline
\end{tabular}
\caption{\ingress{} Results of Reliability of Gratifications}
\label{tab:Ingress-reliability-grats}
\end{table}

Overall, the (modified) Traits sections and (modified) Gratification sections in \pogo{}, \hp{} and \ingress{} surveys are all shown to be valid and reliable. 
Therefore, we can further investigate the casual relationships and correlations between variables based on the data acquired through these instruments. 

\section{Results}
\subsection{The casual and hardcore players of \pogo{}, \hp{}, and \ingress{}.}

Based on the statistics provided in Table \ref{tab:casual-hardcore-players}, it can be found that most of the participants tend to play the 3 games casually as 67.6\% of \pogo{} players, 72.8\% of \hp{} players and 60.0\% of the \ingress{} players describes themselves as casual players. 
Although hardcore players generally account for a relatively small part of the whole population in each game, \ingress{} has the most hardcore players among all 3 games, reaching a 40\%. 
In addition, the demographic information of casual and hardcore players in our survey is for the most part the same. 
This means that both casual and hardcore players in the 3 games are mainly male, employed, well-educated and from English-speaking countries as well as European countries.

That being said, \pogo{} and \hp{} players are in average under 36 years old, while \ingress{} players are in average over 36 years old. 
However, we find that in comparison to casual players, hardcore players of all the 3 games tend to play more often and play for longer time. 
In detail, 84.7\% of the \pogo{} hardcore players reported that they would play the game many times a day and 53.9\% of them would play for more than 30 minutes. 
In contrast, only over half of the casual players (54.8\%) would play \pogo{} multiple times every day and most of them (60.2\%) would play for less than half an hour. 
Regarding \hp{}, although many casual players (62.7\%) also play the game many times a day, more hardcore players (85.8\%) would play at the same level of frequency. 

More importantly, over half of the hardcore players (52.8\%) would play for more than half an hour in each game session while only around a quarter (26.8\%) of the casual players would spend the same amount of time. 
The differences in game frequency and time between casual and hardcore players are more obvious in \ingress{}, in which the percentage of hardcore players who play multiple times per day (76.2\%) is over twice that among casual players (35.7\%). 
Moreover, over 75\% of the \ingress{} hardcore players would spend more than 30 minutes on each game session while the rest a quarter of the hardcore players of \ingress{} would spend relatively shorter time within half an hour. 
In contrast, nearly half of the casual players of \ingress{} (49.6\%) would spend less than 30 minutes on each game session while the other half would spend more than half an hour every time they play \ingress{}.

\begin{table}[htpb]
\begin{tabular}{|r|l|l|l|l|l|l|}
\hline
\multicolumn{1}{|l|}{\multirow{2}{*}{\textbf{Characteristics}}} & \multicolumn{2}{c|}{\textbf{\pogo{} Players}} & \multicolumn{2}{c|}{\textbf{HPWU players}} & \multicolumn{2}{c|}{\textbf{\ingress{} players}} \\ \cline{2-7} 
\multicolumn{1}{|l|}{} & \textbf{Casual} & \textbf{Hardcore} & \textbf{Casual} & \textbf{Hardcore} & \textbf{Casual} & \textbf{Hardcore} \\ \hline
\multicolumn{1}{|l|}{\textbf{Statistics}} &  &  &  &  &  &  \\ \hline
Number & 724 & 347 & 769 & 288 & 157 & 105 \\ \hline
Percentage & 67.60\% & 32.40\% & 72.80\% & 27.20\% & 60.00\% & 40.00\% \\ \hline
\multicolumn{1}{|l|}{\textbf{Gender}} &  &  &  &  &  &  \\ \hline
Male & 518 (71.5\%) & 279 (80.4\%) & 381 (49.5\%) & 159 (55.2\%) & 130 (82.8\%) & 81 (77.1\%) \\ \hline
Female & 191 (26.4\%) & 61 (17.6\%) & 378 (49.2\%) & 120 (41.7\%) & 25 (15.9\%) & 22 (21.0\%) \\ \hline
Non-binary & 15 (2.1\%) & 7 (2.0\%) & 10 (1.3\%) & 9 (3.1\%) & 2 (1.3\%) & 2 (1.9\%) \\ \hline
\multicolumn{1}{|l|}{\textbf{Age}} &  &  &  &  &  &  \\ \hline
$\leq$ 20 years & 162 (22.4\%) & 57 (16.4\%) & 30 (3.9\%) & 11 (3.8\%) & 11 (7.0\%) & 9 (8.6\%) \\ \hline
21-35 years & 462 (63.8\%) & 237 (68.3\%) & 527 (68.5\%) & 196 (68.1) & 68 (43.3\%) & 30 (28.6\%) \\ \hline
$\geq$ 36 years & 100 (13.8\%) & 53 (15.3\%) & 212 (27.6\%) & 81 (28.1\%) & 78 (49.7\%) & 66 (62.8\%) \\ \hline
Occupation &  &  &  &  &  &  \\ \hline
Employed & 429 (59.3\%) & 213 (61.4\%) & 581 (75.6\%) & 232 (80.6\%) & 121 (77.1\%) & 80 (76.2\%) \\ \hline
Unemployed & 62 (8.6\%) & 33 (9.5\%) & 72 (9.3\%) & 22 (7.6\%) & 14 (8.9\%) & 17 (16.3\%) \\ \hline
\multicolumn{1}{|l|}{\textbf{Student}} & 233 (32.1\%) & 101 (29.1\%) & 116 (15.1\%) & 34 (11.8\%) & 22 (14.0\%) & 10 (9.5\%) \\ \hline
Education &  &  &  &  &  &  \\ \hline
$\geq$University degree & 364 (50.3\%) & 178 (51.3\%) & 500 (65.0\%) & 181 (62.8\%) & 88 (56.1\%) & 45 (42.9\%) \\ \hline
College degree & 88 (12.2\%) & 41 (11.8\%) & 124 (16.1\%) & 49 (17.0\%) & 20 (12.7\%) & 21 (20.0\%) \\ \hline
Vocational degree & 28 (3.9\%) & 16 (4.6\%) & 22 (2.9\%) & 12 (4.2\%) & 3 (1.9\%) & 2 (1.9\%) \\ \hline
High school or lower & 244 (33.6\%) & 112 (32.3\%) & 123 (16.0\%) & 46 (16.0\%) & 46 (29.3\%) & 37 (35.2\%) \\ \hline
\multicolumn{1}{|l|}{\textbf{Countries}} &  &  &  &  &  &  \\ \hline
United States & 337 (53.5\%) & 161 (46.4\%) & 391 (50.8\%) & 143 (49.7\%) & 73 (46.5\%) & 53 (50.5\%) \\ \hline
United Kingdom & 89 (12.3\%) & 34 (9.8\%) & 75 (9.8\%) & 18 (6.3\%) & 25 (15.9\%) & 19 (18.1\%) \\ \hline
Germany & 36 (5.0\%) & 29 (8.4\%) & 48 (6.2\%) & 8 (2.8\%) & 11 (7.0\%) & 6 (5.7\%) \\ \hline
Canada & 42 (5.8\%) & 14 (4.0\%) & 35 (4.6\%) & 25 (8.7\%) & 8 (5.1\%) & 3 (2.9\%) \\ \hline
Australia & 27 (3.7\%) & 13 (3.7\%) & 14 (1.8\%) & 6 (2.1\%) & 2 (1.3\%) & 1 (1.0\%) \\ \hline
Netherlands & 15 (2.1\%) & 9 (2.6\%) & 20 (2.6\%) & 5 (1.7\%) & 5 (3.2\%) & 0 (0\%) \\ \hline
Poland & 14 (1.9\%) & 4 (1.4\%) & 12 (1.6\%) & 2 (0.7\%) & 1 (0.6\%) & 0 (0\%) \\ \hline
Italy & 7 (1.0\%) & 10 (2.9\%) & 10 (1.3\%) & 2 (0.7\%) & 1 (0.6\%) & 0 (0\%) \\ \hline
France & 11 (1.5\%) & 5 (1.4\%) & 8 (1.0\%) & 5 (1.7\%) & 1 (0.6\%) & 2 (1.9\%) \\ \hline
Finland & 11 (1.5\%) & 5 (1.4\%) & 14 (1.8\%) & 3 (1.0\%) & 2 (1.3\%) & 2 (1.9\%) \\ \hline
Others & 135 (18.6\%) & 63 (18.1\%) & 142 (18.5\%) & 51 (17.8\%) & 28 (17.8\%) & 19 (18.1\%) \\ \hline
\multicolumn{1}{|l|}{\textbf{Play Frequency}} &  &  &  &  &  &  \\ \hline
Many times a day & 397 (54.8\%) & 294 (84.7\%) & 482 (62.7\%) & 247 (85.8\%) & 56 (35.7\%) & 80 (76.2\%) \\ \hline
Once a day & 218 (30.1\%) & 39 (11.2\%) & 213 (27.7\%) & 40 (13.9\%) & 47 (29.9\%) & 15 (14.3\%) \\ \hline
Few times a week & 89    12.3\%) & 13 (3.7\%) & 53 (6.9\%) & 0 (0.0\%) & 39 (24.8\%) & 9 (8.5\%) \\ \hline
Once a week & 4 (0.6\%) & 1 (0.3\%) & 6 (0.8\%) & 0 (0.0\%) & 10 (6.4\%) & 1 (1\%) \\ \hline
Few times a month & 10 (1.4\%) & 0 (0.0\%) & 6 (0.8\%) & 0 (0.0\%) & 4 (2.5\%) & 0 (0\%) \\ \hline
Rarely & 6 (0.8) & 0 (0.0\%) & 9 (1.2\%) & 1 (0.3\%) & 1 (0.6\%) & 0 (0\%) \\ \hline
\multicolumn{1}{|l|}{\textbf{Game session}} &  &  &  &  &  &  \\ \hline
$\leq$15 mins & 211 (29.1\%) & 61 (17.6\%) & 298 (38.8\%) & 19 (17.0\%) & 23 (14.6\%) & 11 (10.5\%) \\ \hline
16-30 mins & 225 (31.1\%) & 99 (28.5\%) & 265 (34.5\%) & 87 (30.2\%) & 55 (35.0\%) & 15 (14.3\%) \\ \hline
31-45 mins & 101 (14.0\%) & 41 (11.8\%) & 97 (12.6\%) & 45 (15.6\%) & 22 (14.0\%) & 12 (11.4\%) \\ \hline
46-60 mins & 78 (10.8\%) & 36 (10.4\%) & 49 (6.4\%) & 45 (15.6\%) & 26 (16.7\%) & 18 (17.0\%) \\ \hline
\textgreater{}60 mins & 109 (15.0\%) & 110 (31.7\%) & 60 (7.8\%) & 62 (21.6\%) & 31 (19.7\%) & 49 (46.7\%) \\ \hline
\end{tabular}
\caption{The casual and hardcore players of \pogo{}, \hp{}, and \ingress{}}
\label{tab:casual-hardcore-players}
\end{table}

\subsection{The traits of casual and hardcore players of \pogo{}, \hp{}, and \ingress{}}

The results of this section can be found summarized in Table \ref{tab:poke-casual-hardcore-traitgrat}.
In order to better understand the results, there are several items that need to be mentioned before looking into the detailed data:

\begin{itemize}
    \item Except for Nostalgia dimension, the items in each dimension are scored as ``1 - Strongly agree, 2 - Agree, 3 - Somewhat agree, 4 – Neither agree nor disagree, 5 – Somewhat disagree, 6 – Disagree, 7 – Strongly disagree''. Therefore, a lower average score in each dimension represents a greater acknowledgement of participants to the trait or gratification specified in this dimension while a higher score indicate that participants are less likely to have a certain trait or gratification.
    \item Items in Nostalgia (e.g. ``I used to read Harry Potter books.'') are binary, in which 1 stands for ``Yes'' and 2 stands for ``No''. Therefore, a mean close to 1 indicates that participants tend to have nostalgia in the game while a mean close to 2 indicates that they do not have this gratification.
    \item Independent sample T-test is conducted in order to investigate the differences in certain traits and gratifications between casual and hardcore players. The confidence interval is set up to 95\%. Once the significance in certain dimension is less than 0.05, we reckon that the difference in this dimension between casual and hardcore players is statistically significant.  
\end{itemize}

\subsubsection{The traits and gratifications of casual and hardcore players of \pogo{}} \label{tg-pogo}
Again, please note that the results relevant to this section can be seen in Table \ref{tab:poke-casual-hardcore-traitgrat}.

\paragraph{Aesthetic - Traits:} All players of \pogo{} like the aesthetic experiences while casual players are more likely to be aesthetic-oriented. 
This is because that both their mean values are close to 2 which indicates that they “agree” with the items in this dimension such as “I like to spend some time exploring the game world.”
However, in comparison to hardcore player, casual players scored significantly less than hardcore players with p value less than 0.05 at 0.035. 
This means they are more likely to enjoy the game world, sounds, graphics and art-style in \pogo{}.

\paragraph{Immersion - Traits:} All players prefer complex stories and narratives in \pogo{} without significant differences between the casual and hardcore players. 
This is because that (1) their mean values which are close to 2 indicate that they generally “agree” with the items in Immersion Orientation dimension that include items such as “Story is important to me when I play games.”, and (2) that the p value at 0.057 indicates the differences between these two types of player are not statistically significant. 

\paragraph{Goal - Traits:} All players would like to complete requests and tasks in \pogo{} yet hardcore players tend to be more goal oriented in comparison to casual players. 
This is because that both the average score of hardcore players and casual players are around 3, which suggests that they, to some extent, agree with the items in Goal Orientation dimension. 
One typical example can be “I usually do care if I do not complete all optional parts of a game.” 
Moreover, a p value less than 0.05 at 0.000 suggests that hardcore players who scored less in this dimension are more likely to finish more tasks with higher completeness.

\paragraph{Social - Traits:} All players prefer to play \pogo{} with others online or in the same places while hardcore players are more socially oriented.
This is also based on the facts that (1) both casual and hardcore players scored around 3 which indicates that they somewhat agree that they like to, for example, “play with other people”, and (2) that hardcore players scored significantly less than casual players. 
When taken together, the results suggest that players are more likely to interact with others during game sessions. 

\paragraph{Action - Traits:} When \pogo{} becomes a bit more difficult, all players like the challenges. 
Hardcore players tend to enjoy them more compared to casual players. This is because that the average scores of both casual players (mean = 3.1492) and hardcore players (mean = 2.7764) indicate that they at least partially agree with items in Action Orientation dimension, given 3 means “somewhat agree”. 
One typical example of these items can be that “I like it when goals are hard to achieve in games”. 
Moreover, the p value less than 0.05 reveals that the difference in the preference of hardcore players for in-game difficulties is significantly higher than casual players, given that lower mean scores indicate greater acknowledgements to the dimensions.

\paragraph{Socialization - Gratification:}  Hardcore players tend to be gratified as they can make new friends outdoors and maintain relationships with others through \pogo{} while casual players are more likely to have a neutral attitude towards these social activities. 
The average score of hardcore players (mean = 3.3833) shows that they to some extent agree that \pogo{} would enable them to “be part of a group” or “meet strangers outdoors” while the average score of casual players (mean = 4.2778) suggests that they generally “neither agree nor disagree” with these items in Socialization dimension. 
Besides, the p value under 0.05 also indicates that the differences in socialization-related gratification between hardcore players and casual players is statistically significant.

\paragraph{Enjoyment - Gratification:}  All players are gratified by the enjoyment \pogo{} provides while hardcore players gain more enjoyment from playing the game. 
This is because that both casual players and hardcore players scored averagely around 2, which indicates that they generally agree with the items in Enjoyment dimension. 
A typical item is “I play \pogo{} because it is fun”.  
In addition, given hardcore players (mean = 2.2032) scored less than casual players (mean = 2.4848) and the p value is less than 0.05, it can be safely said (95\% sure due to the confidence interval) that hardcore players gain significantly more pleasures in \pogo{} compared to casual players.

\paragraph{Trendiness - Gratification:} Given the mean scores of both casual players (mean = 5.9517) and hardcore players (mean = 5.7041) are close to 6 which stands for “Disagree” in our scale, the results suggest that all players generally do not reckon that \pogo{} would enable them to “look cool/trendy/stylish”. 
Despite that trendiness is not an important gratification, hardcore players are more likely to be gratified with the popularity of the game as they scored relatively lower than casual players in this dimension and the p value equals to 0.005 indicates the difference is statistically significant. 

\paragraph{Competition - Gratification:} All players are gratified by the competition in \pogo{} yet hardcore players tend to be more satisfied when they perform better or beat others during the game.  
This is because that the hardcore players (mean = 2.3317) generally agree the items in Competition dimension such as “I like to prove that I am one of the best players.”, while casual players would partially agree with these items given their mean is 2.7276 which is more close to “3 – somewhat agree”. 
Furthermore, the p value less than 0.05 also indicates that the perceived gratification from in-game competitions is significantly different between hardcore players and casual players. 

\paragraph{Challenge - Gratification:} In-game challenges do not result in gratification among casual players while hardcore players think that being “tested” is fine and are more gratified after overcoming the difficulties. 
This result is first reflected in the data that casual player averagely scored up to 4.8020 which is close to “5 – somewhat disagree” in Challenge dimension. 
This means items such as “I feel proud when I master an aspect of \pogo{}” may not be the case among casual players. 
In contrast, hardcore players tend to have a neutral attitude towards in-game challenges given their mean score in this dimension, 4.0038, is close to “4 – neither agree nor disagree”. 
This suggests that at least some items or some situations illustrated in the items correctly describe the gratification that hardcore players gained from challenges so that the overall score reached a relatively balanced point. 
In addition, the significant p value (p<0.05) indicates that hardcore players would have significantly more gratifications from challenges in comparison to causal players.

Interestingly, although hardcore players show their favor to in-game difficulties consistently, casual players who are found to prefer challenging games in Traits section now tend to reckon that in-game challenges would not result in gratifications. 
This indicates that the observed personal preferences may not correspondingly lead to expected gameplay experiences. 
Therefore, the relationships between traits and gratifications need to be further studied and game designers should consider both aspects while developing games.

\paragraph{Nostalgia - Gratification:} Both casual players and hardcore players of \pogo{} are gratified by nostalgia while there is no significant difference between the two groups. 
Given that the items in nostalgia (e.g. “I used to read Pok{\'e}mon manga/books.”) are binary questions in which 1 stands for Yes while 2 stands for No, the average scores of both casual players (mean = 1.2175) and hardcore players (mean = 1.2341) indicate that all players are likely to get to know Pok{\'e}mon before the game and therefore get gratified by a yearning to relive or return to a past period during game sessions. 
Besides, p value equals to 0.451 suggests that there is no significant difference in Nostalgia dimension between casual and hardcore players.

\begin{table}[htpb]
\begin{tabular}{|r|l|l|l|l|l|l|l|}
\hline
\multicolumn{1}{|c|}{\multirow{2}{*}{\textbf{\pogo{}}}} & \multicolumn{2}{c|}{\textbf{Casual (724)}} & \multicolumn{2}{c|}{\textbf{Hardcore (347)}} & \multicolumn{1}{c|}{\multirow{2}{*}{\textbf{Dif}}} & \multicolumn{1}{c|}{\multirow{2}{*}{\textbf{Sig (p)}}} & \multicolumn{1}{c|}{\multirow{2}{*}{\textbf{t}}} \\ \cline{2-5}
\multicolumn{1}{|c|}{} & \multicolumn{1}{c|}{\textbf{Mean}} & \multicolumn{1}{c|}{\textbf{SD}} & \multicolumn{1}{c|}{\textbf{Mean}} & \multicolumn{1}{c|}{\textbf{SD}} & \multicolumn{1}{c|}{} & \multicolumn{1}{c|}{} & \multicolumn{1}{c|}{} \\ \hline
\multicolumn{1}{|l|}{\textbf{Traits}} &  &  &  &  &  &  &  \\ \hline
Aesthetic Orientation & 2.1019 & 0.83839 & 2.2184 & 0.8601 & -0.11651 & 0.035 & -2.111 \\ \hline
Immersion Orientation & 2.3719 & 1.19351 & 2.5291 & 1.30297 & -0.15784 & 0.057 & -1.906 \\ \hline
Goal Orientation & 3.0072 & 1.03518 & 2.7493 & 1.0093 & 0.2579 & 0 & 3.847 \\ \hline
Social Orientation & 3.1925 & 1.53376 & 2.8571 & 1.42207 & 0.33548 & 0 & 3.521 \\ \hline
Action Orientation & 3.1492 & 1.24426 & 2.7764 & 1.20529 & 0.3728 & 0 & 4.635 \\ \hline
\multicolumn{1}{|l|}{\textbf{Gratifications}} &  &  &  &  &  &  &  \\ \hline
Socialization & 4.2778 & 1.47517 & 3.3833 & 1.34148 & 0.89451 & 0 & 9.883 \\ \hline
Enjoyment & 2.4848 & 0.98147 & 2.2032 & 0.95214 & 0.28164 & 0 & 4.437 \\ \hline
Trendiness & 5.9517 & 1.21866 & 5.7041 & 1.42108 & 0.24753 & 0.005 & 2.79 \\ \hline
Competition & 2.7276 & 1.02101 & 2.3317 & 0.89719 & 0.39614 & 0 & 6.461 \\ \hline
Challenge & 4.802 & 1.38071 & 4.0038 & 1.4701 & 0.79818 & 0 & 8.668 \\ \hline
Nostalgia & 1.2175 & 0.33417 & 1.2341 & 0.34492 & -0.01661 & 0.451 & -0.753 \\ \hline
\end{tabular}
\caption{\pogo{} Traits and Gratifications values for Casual and Hardcore Players}
\label{tab:poke-casual-hardcore-traitgrat}
\end{table}

\subsubsection{The traits and gratifications of casual and hardcore players of \hp{}} \label{tg-hp} 
Considering that (1) the results in this section are also analyzed based on the mean values and p values, and (2) the dimensions as well as the involved items in \hp{} are similar to those in \pogo{} section, we decide to simply describe the results without detailed explanations for the sake of brevity.
The results relevant to this section can be seen in Table \ref{tab:hp-casual-hardcore-traitgrat}.

\paragraph{Aesthetic - Traits:} Both casual players (mean = 2.2224) and hardcore players (mean = 2.1278) prefer aesthetic experiences to explore the game world and appreciate the game’s graphics, sound, and art style, given 2 stands for “Agree” in the scale of the questionnaire. 
No significant difference is detected between the two groups given p value is more than 0.05 at 0.098.

\paragraph{Immersion - Traits:} Both casual players (mean = 2.5246) and hardcore players (mean = 2.1278) like the complex stories and narratives in Harry Potter United Wizards given 2 stands for “Agree” in the scale of the questionnaire. 
No significant difference is detected between the two groups as p value is more than 0.05 at 0.081.

\paragraph{Goal - Traits:} Both casual players (mean = 2.7508) and hardcore players (mean = 2.3826) have the preferences for gameplay that involves completing quests or tasks, collecting digital objects, or similar experiences. 
Hardcore players are more likely to be goal oriented as they scored less in this dimension and the p value is significant with value less than 0.05. 

\paragraph{Social - Traits:} Both casual players (mean = 3.5732) and hardcore players (mean = 3.1750) partially admit that they prefer to play together with others online or in the same space, given 3 stands for “Somewhat agree” in the scale of the questionnaire. 
Hardcore players are more likely to play with others as they scored less in this section and the p value is less than 0.05.

\paragraph{Action - Traits:} Both casual players (mean = 3.3599) and hardcore players (mean = 2.8924) to some extends agree that they like challenging and fast-paced gameplays, given 3 stands for “Somewhat agree” in the scale of the questionnaire. 
Hardcore players are more likely to enjoy the challenges, for example playing with more difficult setting, as they scored less in this section and the p value is less than 0.05.

\paragraph{Socialization – Gratification:} Casual players (mean = 5.0683) are not gratified by the social activities (e.g. meet others outdoor, maintain relationships with others) while hardcore players (mean = 4.4983) tend to have a neutral attitudes towards these activities, given that 5 stands for “Somewhat disagree” and 4 stands for “neither agree nor disagree” in the scale of the questionnaire. 
However, hardcore players tend to be more satisfied with the interactions with others through \hp{} compared to casual players as they scored less in this dimension with a p value less than 0.05 which shows the significance of the difference.

\paragraph{Enjoyment – Gratification:} Casual players (mean = 2.8443) tend to partially agree that they enjoyment is a gratification they derive from \hp{} as their mean score is more close to “3 – somewhat agree” while hardcore players would reckon enjoyment as one of their gratification to a greater extend as their mean score is more close to “2 - Agree” and the p value less than 0.05 also indicate that this difference is significant. 

\paragraph{Trendiness – Gratification:} Both casual players (mean = 6.0251) and hardcore players (mean = 5.9514) would not be gratified by the trendiness of \hp{}, given a mean score at 6 indicate that participants “disagree” with the items involved in certain dimension. No significant difference between the two types of players is detected given the p value is larger than 0.05 at 0.396.

\paragraph{Competition – Gratification:} Casual players (mean = 2.8637) tend to partially agree that they are gratified by the game-related competitions in \hp{} as their mean score is more close to “3 – somewhat agree” while hardcore players, whose mean score is more close to “2 - Agree”, would have more gratifications if they can perform better than others in the game. Besides, the p value less than 0.05 also indicate that this difference is significant.

\paragraph{Challenge – Gratification:} Hardcore players (mean = 3.8194) tend to have a neutral attitude towards the gratification gained from in-game challenges, given 4 stands for “neither agree nor disagree” in the scale of the questionnaire. In addition, hardcore players would also be more satisfied once they conquer the difficulties (p<0.05) in comparison to casual players who generally “disagree” (mean = 5.9514) that they can be gratified by the in-game challenges. 

\paragraph{Nostalgia – Gratification:} Given the mean scores of both casual players (mean = 1.2263) and hardcore players (mean = 1.1910) are close to 1 which stands for “Yes” in nostalgia-related binary items, all the players are more likely to be gratified by a yearning to relive or return to a past period. However, according to the p value that is less than 0.05, hardcore players of \hp{} tend to gain significantly more gratification from the feeling of nostalgia in comparison to their casual counterparts. 

\begin{table}[htpb]
\begin{tabular}{|r|l|l|l|l|l|l|l|}
\hline
\multicolumn{1}{|c|}{\multirow{2}{*}{\hp{}}} & \multicolumn{2}{c|}{\textbf{Casual (769)}} & \multicolumn{2}{c|}{\textbf{Hardcore (228)}} & \multicolumn{1}{c|}{\multirow{2}{*}{\textbf{Difference}}} & \multicolumn{1}{c|}{\multirow{2}{*}{\textbf{Sig (p)}}} & \multicolumn{1}{c|}{\multirow{2}{*}{\textbf{t}}} \\ \cline{2-5}
\multicolumn{1}{|c|}{} & \multicolumn{1}{c|}{\textbf{Mean}} & \multicolumn{1}{c|}{\textbf{SD}} & \multicolumn{1}{c|}{\textbf{Mean}} & \multicolumn{1}{c|}{\textbf{SD}} & \multicolumn{1}{c|}{} & \multicolumn{1}{c|}{} & \multicolumn{1}{c|}{} \\ \hline
\multicolumn{1}{|l|}{\textbf{Traits}} &  &  &  &  &  &  &  \\ \hline
Aesthetic Orientation & 2.2224 & 0.83099 & 2.1278 & 0.81488 & 0.09459 & 0.098 & 1.656 \\ \hline
Immersion Orientation & 2.5246 & 1.30077 & 2.3681 & 1.29392 & 1.5652 & 0.081 & 1.744 \\ \hline
Goal Orientation & 2.7508 & 1.03899 & 2.3826 & 1.10462 & 0.36821 & 0 & 5.041 \\ \hline
Social Orientation & 3.5732 & 1.59722 & 3.175 & 1.57522 & 0.39821 & 0 & 3.622 \\ \hline
Action Orientation & 3.3599 & 1.24719 & 2.8924 & 1.24836 & 0.46759 & 0 & 5.426 \\ \hline
\multicolumn{1}{|l|}{\textbf{Gratification}} &  &  &  &  &  &  &  \\ \hline
Socialization & 5.0683 & 1.34598 & 4.4983 & 1.47386 & 0.57001 & 0 & 5.729 \\ \hline
Enjoyment & 2.8443 & 1.06558 & 2.3498 & 1.07403 & 0.49445 & 0 & 6.702 \\ \hline
Trendiness & 6.0251 & 1.18668 & 5.9514 & 1.28119 & 0.07375 & 0.396 & 0.88 \\ \hline
Competition & 2.8637 & 1.0466 & 2.1847 & 0.94918 & 0.679 & 0 & 10.063 \\ \hline
Challenge & 4.6808 & 1.2903 & 3.8194 & 1.33178 & 0.86131 & 0 & 9.619 \\ \hline
Nostalgia & 1.2263 & 0.24607 & 1.191 & 0.22377 & 0.0353 & 0.034 & 2.127 \\ \hline
\end{tabular}
\caption{ \hp{} Traits and Gratifications values for Casual and Hardcore Players}
\label{tab:hp-casual-hardcore-traitgrat}
\end{table}

\subsubsection{The traits and gratifications of casual and hardcore players of \ingress{}} \label{tg-ingress}

Based on the data in Table \ref{tab:ingress-casual-hardcore-traitgrat}, it can be observed both casual players and hardcore players tend to partially agree with the items in all the dimensions in Traits section as their mean scores of each dimension are all close to 3 which stands for “Somewhat agree” in the scale of the questionnaire. 
Therefore, it can be summarized that both casual players and hardcore players of \ingress{} to some extent have preferences for: 

\begin{itemize}
    \item Exploring the in-game landscape, graphics, sound and art style (Aesthetic Orientation) 
    \item Complex stories and narratives in \ingress{} (Immersion Orientation) 
    \item Gameplay that involves completing quests or tasks, collecting digital objects, or similar experiences (Goal Orientation)
    \item Playing together with others online or in the same space (Social Orientation)
    \item Challenging and fast-paced gameplays (Action Orientation)
\end{itemize}

Notably, hardcore players of \ingress{} tend to be more goal oriented and action oriented compared to casual players as the p values in these two dimensions are less than 0.05, which indicate that the differences between the type of players are significant.
Since there is no common pattern in the gratifications of casual and hardcore players of \ingress{}, we decide to report the results separately as we did in the previous sections. In detail:

\paragraph{Socialization – Gratification:} Casual players (mean = 3.8654) tend to have a neutral attitude towards socializing through \ingress{} (e.g. meet others outdoor, maintain relationships with others) while hardcore players (mean = 3.0429) is more gratified by these activities, given 4 stands for “neither agree nor disagree” while 3 stands for “somewhat agree”. 
In addition, p value less than 0.05 also suggests that hardcore players would have significantly more gratifications from interaction with other players.

\paragraph{Enjoyment - Gratification:} Both casual players (mean = 2.5074) and hardcore players (mean = 2.1460) would at least partially agree that they would gain gratifications from leisure in \ingress{} as 3 stands for “somewhat agree” while 2 stands for “agree” in the scale of the questionnaire. 
Furthermore, in comparison to casual players, hardcore players would be more gratified by the enjoyments in \ingress{} as the p value at 0.004 indicate the difference between the two groups is significant.

\paragraph{Trendiness – Gratification:} Similar to \pogo{} and \hp{}, both casual players (mean = 5.8854) and hardcore players (mean = 5.8603) are not be gratified by the popularity of \ingress{} as their mean values are close to 6, which indicates that the participants tend to simply disagree with the items in Trendiness dimension. 
Also, the p value at 0.875 shows no significant difference between the two groups.

\paragraph{Competition – Gratification:} Both causal players (mean = 2.3439) and hardcore players (mean = 1.8619) would gain gratifications from the in-game competitions in \ingress{}, given both their mean scores are close to “2 – Agree”.  
Moreover, compared to casual players, hardcore players would be more gratified if they beat others in the game as the p value less than 0.05 indicate that the difference in the perception of winning competitions between casual and hardcore players is statistically significant.

\paragraph{Challenge – Gratification:} Given that 5 stands for “Somewhat disagree” while 4 indicates participant “neither agree or disagree” with the items, casual players (mean = 4.5138) are found to partially disagree that in-game challenges results in gratification while hardcore players (mean = 3.8508) would have a neutral attitudes. 
In addition, hardcore player would be more gratified in comparison to casual players when they conquer the in-game challenges as the p value less than 0.05 suggests that this difference is statistically significant.

\begin{table}[htpb]
\begin{tabular}{|r|l|l|l|l|l|l|l|}
\hline
\multicolumn{1}{|c|}{\multirow{2}{*}{\textbf{\ingress{}}}} & \multicolumn{2}{c|}{\textbf{Casual (157)}} & \multicolumn{2}{c|}{\textbf{Hardcore (105)}} & \multicolumn{1}{c|}{\multirow{2}{*}{\textbf{Difference}}} & \multicolumn{1}{c|}{\multirow{2}{*}{\textbf{Sig(p)}}} & \multicolumn{1}{c|}{\multirow{2}{*}{\textbf{t}}} \\ \cline{2-5}
\multicolumn{1}{|c|}{} & \multicolumn{1}{c|}{\textbf{Mean}} & \multicolumn{1}{c|}{\textbf{SD}} & \multicolumn{1}{c|}{\textbf{Mean}} & \multicolumn{1}{c|}{\textbf{SD}} & \multicolumn{1}{c|}{} & \multicolumn{1}{c|}{} & \multicolumn{1}{c|}{} \\ \hline
\multicolumn{1}{|l|}{\textbf{Traits}} &  &  &  &  &  &  &  \\ \hline
Aesthetic Orientation & 2.6497 & 1.03266 & 2.5571 & 1.07208 & .09254 & .485 & .700 \\ \hline
Immersion Orientation & 3.0357 & 1.54364 & 3.3486 & 1.57701 & -.31290 & .112 & -1.594 \\ \hline
Goal Orientation & 3.2191 & 1.18404 & 2.9295 & 1.12146 & .28958 & .049 & 1.981 \\ \hline
Social Orientation & 3.1248 & 1.44123 & 2.8229 & 1.44896 & .30198 & .098 & 1.658 \\ \hline
Action Orientation & 3.0994 & 1.11600 & 2.7352 & 1.15575 & .36412 & .011 & 2.551 \\ \hline
\multicolumn{1}{|l|}{\textbf{Gratification}} &  &  &  &  &  &  &  \\ \hline
Socialization & 3.8654 & 1.32224 & 3.0429 & 1.20144 & .82259 & .000 & 5.116 \\ \hline
Enjoyment & 2.5074 & .95823 & 2.1460 & .99992 & .36140 & .004 & 2.940 \\ \hline
Trendiness & 5.8854 & 1.17894 & 5.8603 & 1.36640 & .02503 & .875 & .153 \\ \hline
Competition & 2.3439 & .89039 & 1.8619 & .70368 & .48204 & .000 & 4.658 \\ \hline
Challenge & 4.5138 & 1.28407 & 3.8508 & 1.32367 & .66301 & .000 & 4.045 \\ \hline
\end{tabular}
\caption{\ingress{} Traits and Gratifications values for Casual and Hardcore Players}
\label{tab:ingress-casual-hardcore-traitgrat}
\end{table}

\subsection{Different traits and gratifications of the casual and hardcore players of \pogo{}, \hp{}, and \ingress{}}
\subsubsection{Different traits and gratifications of casual and hardcore players of \pogo{} and \hp{}}

Since the detailed traits and gratifications of casual and hardcore players of \pogo{} and \hp{} have been reported in the previous sections (see sections  \ref{tg-pogo} and \ref{tg-hp}), we would like to focus on the significant differences (p value less than 0.05) in traits and gratifications between casual and hardcore players of \pogo{} and \hp{}.
The results relevant to this section can be seen in Table \ref{tab:PoGO-HPWU-casual} and Table \ref{tab:PoGO-HPWU-hardcore}.
In detail:

\begin{itemize}
    \item Aesthetic – Traits: Casual players of \pogo{} tend to have more preferences for exploring the game world, graphics, sound and art style in comparison to the casual players of \hp{}. However, there is no significant difference between the hardcore players of both 2 games.
    \item Immersion – Traits: Casual players of \pogo{} are more likely to enjoy the complex stories and narratives in comparison to casual players in \hp{}, while there is no significant difference between the hardcore players in both games.
    \item Goal – Traits: Both the casual players and hardcore players of \pogo{} tend to have less preferences for gameplay that involves completing quests or tasks, collecting digital objects in comparison to the casual players and hardcore players of \hp{}.
    \item Social – Traits: Both the casual players and hardcore players of \pogo{} are more likely to play with others during the game sessions in comparison to the casual players and hardcore players of \hp{}.
    \item Action – Traits: Casual players of \pogo{} tend to prefer challenging and fast-paced gameplays compared to the casual players of \hp{}. However, there is no significant difference between the hardcore players of both games.
    \item Socialization – Gratification: Both the casual players and hardcore players of \pogo{} are more likely to be gratified by the social activities (e.g. meet others outdoor, maintain relationships with others) in comparison to the casual players and hardcore players of \hp{}.
    \item Enjoyment – Gratification: Casual players of \pogo{} tend to be more gratified by the in-game pleasures in comparison to the casual players of \hp{}. There is no significant difference between the hardcore players of both games
    \item Trendiness – Gratification: Hardcore players of \pogo{} are more likely to be gratified by the popularity of the game in comparison to hardcore players of \hp{}. Nonetheless, there is no significant difference between the casual players of both games
    \item Competition – Gratification: Compared to the casual players of \hp{}, the casual players of \pogo{} tend to be more gratified if they win competitions. Interestingly, the hardcore players of \pogo{} tend to be less gratified after winning in-game competitions compared to the hardcore players of \hp{}.
\end{itemize}

\begin{table}[htpb]
\begin{tabular}{|r|c|c|c|c|c|c|c|}
\hline
\multicolumn{1}{|c|}{\textbf{Casual Players}} & \multicolumn{2}{c|}{\textbf{\pogo{} (724)}} & \multicolumn{2}{c|}{\textbf{HPWU (769)}} & \multirow{2}{*}{\textbf{Difference}} & \multirow{2}{*}{\textbf{Sig (p)}} & \multirow{2}{*}{\textbf{t}} \\ \cline{1-5}
\multicolumn{1}{|c|}{\textbf{PoGO-HPWU}} & Mean & SD & Mean & SD &  &  &  \\ \hline
\multicolumn{1}{|l|}{\textbf{Traits}} & \multicolumn{1}{l|}{} & \multicolumn{1}{l|}{} & \multicolumn{1}{l|}{} & \multicolumn{1}{l|}{} & \multicolumn{1}{l|}{} & \multicolumn{1}{l|}{} & \multicolumn{1}{l|}{} \\ \hline
Aesthetic Orientation & 2.1019 & 0.83839 & 2.2224 & 0.83099 & -0.12043 & 0.005 & -2.787 \\ \hline
Immersion Orientation & 2.3719 & 1.19351 & 2.5246 & 1.30077 & -0.15331 & 0.018 & -2.375 \\ \hline
Goal Orientation & 3.0072 & 1.03518 & 2.7508 & 1.03899 & 0.25634 & 0 & 4.773 \\ \hline
Social Orientation & 3.1925 & 1.53376 & 3.5732 & 1.59722 & -0.38067 & 0 & -4.692 \\ \hline
Action Orientation & 3.1492 & 1.24426 & 3.3599 & 1.24719 & -0.21078 & 0.001 & -3.267 \\ \hline
\multicolumn{1}{|l|}{\textbf{Gratification}} &  &  &  &  &  &  &  \\ \hline
Socialization & 4.2778 & 1.47517 & 5.0683 & 1.34598 & -0.79047 & 0 & -10.795 \\ \hline
Enjoyment & 2.4848 & 0.98147 & 2.8443 & 1.06558 & -0.35947 & 0 & -6.785 \\ \hline
Trendiness & 5.9517 & 1.21866 & 6.0251 & 1.18668 & -0.07348 & 0.238 & -1.18 \\ \hline
Competition & 2.7276 & 1.02101 & 2.8637 & 1.0466 & -0.13616 & 0.011 & -2.542 \\ \hline
Challenge & 4.802 & 1.38071 & 4.6808 & 1.2903 & 0.12127 & 0.08 & 1.754 \\ \hline
Nostalgia & 1.2175 & 0.33417 & 1.2263 & 0.24607 & -0.00873 & 0.564 & -0.577 \\ \hline
\end{tabular}
\caption{Casual Players of \pogo{} and \hp{} Traits and Gratifications}
\label{tab:PoGO-HPWU-casual}
\end{table}

\begin{table}[htpb]
\begin{tabular}{|r|l|l|l|l|l|l|l|}
\hline
\multicolumn{1}{|l|}{\textbf{Hardcore Players}} & \multicolumn{2}{l|}{\textbf{\pogo{} (347)}} & \multicolumn{2}{l|}{\textbf{HPWU (288)}} & \multirow{2}{*}{\textbf{Difference}} & \multirow{2}{*}{\textbf{Sig (p)}} & \multirow{2}{*}{\textbf{t}} \\ \cline{1-5}
\multicolumn{1}{|l|}{\textbf{PoGO-HPWU}} & \textbf{Mean} & \textbf{SD} & \textbf{Mean} & \textbf{SD} &  &  &  \\ \hline
\multicolumn{1}{|l|}{\textbf{Traits}} &  &  &  &  &  &  &  \\ \hline
Aesthetic Orientation & 2.2184 & 0.8601 & 2.1278 & 0.81488 & 0.09067 & 0.176 & 1.354 \\ \hline
Immersion Orientation & 2.5291 & 1.30297 & 2.3681 & 1.29392 & 0.16105 & 0.12 & 1.555 \\ \hline
Goal Orientation & 2.7493 & 1.0093 & 2.3826 & 1.10462 & 0.36664 & 0 & 4.366 \\ \hline
Social Orientation & 2.8571 & 1.42207 & 3.175 & 1.57522 & -0.31794 & 0.008 & -2.671 \\ \hline
Action Orientation & 2.7764 & 1.20529 & 2.8924 & 1.24836 & -0.11599 & 0.235 & -1.188 \\ \hline
\multicolumn{1}{|l|}{\textbf{Gratification}} &  &  &  &  &  &  &  \\ \hline
Socialization & 3.3833 & 1.34148 & 4.4983 & 1.47386 & -1.11498 & 0 & -9.883 \\ \hline
Enjoyment & 2.2032 & 0.95214 & 2.3498 & 1.07403 & -0.14666 & 0.072 & -1.803 \\ \hline
Trendiness & 5.7041 & 1.42108 & 5.9514 & 1.28119 & -0.24726 & 0.022 & -2.304 \\ \hline
Competition & 2.3317 & 0.89719 & 2.1847 & 0.94918 & 0.14669 & 0.046 & 1.998 \\ \hline
Challenge & 4.0038 & 1.4701 & 3.8194 & 1.33178 & 0.1844 & 0.099 & 1.669 \\ \hline
Nostalgia & 1.2341 & 0.34492 & 1.191 & 0.22377 & 0.04318 & 0.058 & 1.899 \\ \hline
\end{tabular}
\caption{Hardore Players of \pogo{} and \hp{} Traits and Gratifications}
\label{tab:PoGO-HPWU-hardcore}
\end{table}

\subsubsection{Different traits and gratifications of casual and hardcore players of \pogo{} and \ingress{}}
The results relevant to this section can be found in Table \ref{tab:PoGO-ingress-casual} and Table \ref{tab:PoGO-ingress-hardcore}.

\begin{itemize}
    \item Aesthetic – Traits: Both the casual players and hardcore players of \pogo{} are more likely to prefer to explore the game world, graphics, sound and art style in comparison to the casual and hardcore players of \ingress{}. 
    \item Immersion – Traits: Both the casual players and hardcore players of \pogo{} tend to have more preferences for complex stories and narratives in comparison to the casual players and hardcore players of \ingress{}.
    \item Goal – Traits: Casual players of \pogo{} are more likely to prefer games which involves completing quests or tasks, collecting digital objects compared to the casual players of \ingress{}. However, there is no significant difference between the hardcore players of the two games.
    \item Socialization - Gratification: Both the casual players and hardcore players of \pogo{} are less likely to be gratified by the game-related interactions with others in comparison to the casual players and hardcore players of \ingress{}.
    \item Competition – Gratification: Both the casual players and hardcore players of \pogo{} tend to gain less gratification from in-game competitions compared to the casual players and hardcore players of \ingress{}.
    \item Challenge – Gratification: In comparison to casual players of \ingress{}, the casual players of \pogo{} tend to be less gratified when they conquer the in-game challenges.
\end{itemize}

\begin{table}[htpb]
\begin{tabular}{|r|l|l|l|l|l|l|l|}
\hline
\multicolumn{1}{|c|}{\textbf{Casual Players}} & \multicolumn{2}{c|}{\textbf{\pogo{} (724)}} & \multicolumn{2}{c|}{\textbf{\ingress{} (157)}} & \multicolumn{1}{c|}{\multirow{2}{*}{\textbf{Difference}}} & \multicolumn{1}{c|}{\multirow{2}{*}{\textbf{Sig (p)}}} & \multicolumn{1}{c|}{\multirow{2}{*}{\textbf{t}}} \\ \cline{1-5}
\multicolumn{1}{|c|}{\textbf{PoGO-\ingress{}}} & \multicolumn{1}{c|}{\textbf{Mean}} & \multicolumn{1}{c|}{\textbf{SD}} & \multicolumn{1}{c|}{\textbf{Mean}} & \multicolumn{1}{c|}{\textbf{SD}} & \multicolumn{1}{c|}{} & \multicolumn{1}{c|}{} & \multicolumn{1}{c|}{} \\ \hline
\multicolumn{1}{|l|}{\textbf{Traits}} &  &  &  &  &  &  &  \\ \hline
Aesthetic Orientation & 2.1019 & .83839 & 2.6497 & 1.03266 & -.54775 & .000 & -6.217 \\ \hline
Immersion Orientation & 2.3719 & 1.19351 & 3.0357 & 1.54364 & -.66440 & .000 & -5.074 \\ \hline
Goal Orientation & 3.0072 & 1.03518 & 3.2191 & 1.18404 & -.21193 & .039 & -2.077 \\ \hline
Social Orientation & 3.1925 & 1.53376 & 3.1248 & 1.44123 & -.06770 & .613 & .507 \\ \hline
Action Orientation & 3.1492 & 1.24426 & 3.0994 & 1.11600 & .04981 & .620 & .496 \\ \hline
\multicolumn{1}{|l|}{\textbf{Gratification}} &  &  &  &  &  &  &  \\ \hline
Socialization & 4.2778 & 1.47517 & 3.8654 & 1.32224 & .41235 & .001 & 3.232 \\ \hline
Enjoyment & 2.4848 & .98147 & 2.5074 & .95823 & -.02262 & .793 & -.263 \\ \hline
Trendiness & 5.9517 & 1.21866 & 5.8854 & 1.17894 & .06631 & .534 & .622 \\ \hline
Competition & 2.7276 & 1.02101 & 2.3439 & .89039 & .38361 & .000 & 4.361 \\ \hline
Challenge & 4.8020 & 1.38071 & 4.5138 & 1.28407 & .28823 & .017 & 2.400 \\ \hline
\end{tabular}
\caption{Casual Players of \pogo{} and \ingress{} Traits and Gratifications}
\label{tab:PoGO-ingress-casual}
\end{table}

\begin{table}[htpb]
\begin{tabular}{|r|l|l|l|l|l|l|l|}
\hline
\multicolumn{1}{|l|}{\textbf{Hardcore Players}} & \multicolumn{2}{l|}{\textbf{\pogo{} (347)}} & \multicolumn{2}{l|}{\textbf{\ingress{} (105)}} & \multirow{2}{*}{\textbf{Difference}} & \multirow{2}{*}{\textbf{Sig (p)}} & \multirow{2}{*}{\textbf{t}} \\ \cline{1-5}
\multicolumn{1}{|l|}{\textbf{PoGO-\ingress{}}} & \textbf{Mean} & \textbf{SD} & \textbf{Mean} & \textbf{SD} &  &  &  \\ \hline
\multicolumn{1}{|l|}{\textbf{Traits}} &  &  &  &  &  &  &  \\ \hline
Aesthetic Orientation & 2.2184 & .86010 & 2.5571 & 1.07208 & -.33870 & .004 & -2.962 \\ \hline
Immersion Orientation & 2.5291 & 1.30297 & 3.3486 & 1.57701 & -.81946 & .000 & -4.847 \\ \hline
Goal Orientation & 2.7493 & 1.00930 & 2.9295 & 1.12146 & -.18024 & .119 & -1.562 \\ \hline
Social Orientation & 2.8571 & 1.42207 & 2.8229 & 1.44896 & .03420 & .830 & .215 \\ \hline
Action Orientation & 2.7764 & 1.20529 & 2.7352 & 1.15575 & .04113 & .757 & .309 \\ \hline
\multicolumn{1}{|l|}{\textbf{Gratification}} &  &  &  &  &  &  &  \\ \hline
Socialization & 3.3833 & 1.34148 & 3.0429 & 1.20144 & .34043 & .020 & 2.332 \\ \hline
Enjoyment & 2.2032 & .95214 & 2.1460 & .99992 & .05714 & .605 & .519 \\ \hline
Trendiness & 5.7041 & 1.42108 & 5.8603 & 1.36640 & -.15619 & .320 & -.955 \\ \hline
Competition & 2.3317 & .89719 & 1.8619 & .70368 & .46951 & .000 & 4.922 \\ \hline
Challenge & 4.0038 & 1.47010 & 3.8508 & 1.32367 & .15305 & .340 & .956 \\ \hline
\end{tabular}
\caption{Hardcore Players of \pogo{} and \ingress{} Traits and Gratifications}
\label{tab:PoGO-ingress-hardcore}
\end{table}

\subsubsection{Different traits and gratifications of casual and hardcore players of \hp{} and \ingress{}}
The results relevant to this section can be found in Table \ref{tab:HPWU-ingress-casual} and Table \ref{tab:HPWU-ingress-hardcore}.

\begin{itemize}
    \item Aesthetic - Traits: Both the casual players and hardcore players of \hp{} are more likely to prefer to explore the game world, graphics, sound and art style in comparison to the casual and hardcore players of \ingress{}.
    \item Immersion - Traits: Both the casual players and hardcore players of \hp{} tend to have more preferences for complex stories and narratives in comparison to the casual players and hardcore players of \ingress{}.
    \item Goal - Traits: Both the casual players and hardcore players of \hp{} are likely to refer games which involves completing quests or tasks, collecting digital objects compared to the casual players of \ingress{}. 
    \item Social – Traits: Compared to the casual players of \ingress{}, the casual players of \hp{} tend to enjoy more about playing with others during the game sessions. With regard to hardcore players of these two games, however, it is the hardcore player of \ingress{} who are more likely to be social oriented.
    \item Action – Traits: In comparison to casual players of \ingress{}, the casual players of \hp{} are less likely to prefer challenging and fast-paced gameplays.
    \item Socialization – Gratification: Both the casual players and hardcore players of \hp{} would be less gratified by the game-related social activities (e.g. meet others outdoor, maintain relationships with others) in comparison to the casual players and hardcore players of \ingress{}. 
    \item Enjoyment – Gratification: Casual players of \hp{} would be less gratified by the pleasures provided in the game in comparison to the casual players of \ingress{}. However, there is no significant difference between hardcore players of these two games.
    \item Competition – Gratification: Both the casual players and hardcore players of \hp{} would gain less gratifications from in-game competitions compared to the casual and hardcore players of \ingress{}.
\end{itemize}

\begin{table}[htpb]
\begin{tabular}{|r|l|l|l|l|l|l|l|}
\hline
\multicolumn{1}{|l|}{\textbf{Casual Players}} & \multicolumn{2}{l|}{\textbf{HPWU (769)}} & \multicolumn{2}{l|}{\textbf{\ingress{} (157)}} & \multirow{2}{*}{\textbf{Difference}} & \multirow{2}{*}{\textbf{Sig (p)}} & \multirow{2}{*}{\textbf{t}} \\ \cline{1-5}
\multicolumn{1}{|l|}{\textbf{HPWU-\ingress{}}} & \textbf{Mean} & \textbf{SD} & \textbf{Mean} & \textbf{SD} &  &  &  \\ \hline
\multicolumn{1}{|l|}{\textbf{Traits}} &  &  &  &  &  &  &  \\ \hline
Aesthetic Orientation & 2.2224 & .83099 & 2.6497 & 1.03266 & -.42731 & .000 & -4.873 \\ \hline
Immersion Orientation & 2.5246 & 1.30077 & 3.0357 & 1.54364 & -.51109 & .000 & -3.877 \\ \hline
Goal Orientation & 2.7508 & 1.03899 & 3.2191 & 1.18404 & -.46826 & .000 & -4.606 \\ \hline
Social Orientation & 3.5732 & 1.59722 & 3.1248 & 1.44123 & -.44837 & .001 & 3.257 \\ \hline
Action Orientation & 3.3599 & 1.24719 & 3.0994 & 1.11600 & .26058 & .015 & 2.612 \\ \hline
\multicolumn{1}{|l|}{\textbf{Gratification}} &  &  &  &  &  &  &  \\ \hline
Socialization & 5.0683 & 1.34598 & 3.8654 & 1.32224 & 1.20282 & .000 & 10.234 \\ \hline
Enjoyment & 2.8443 & 1.06558 & 2.5074 & .95823 & .33685 & .000 & 3.669 \\ \hline
Trendiness & 6.0251 & 1.18668 & 5.8854 & 1.17894 & .13979 & .178 & 1.347 \\ \hline
Competition & 2.8637 & 1.04660 & 2.3439 & .89039 & .51977 & .000 & 6.460 \\ \hline
Challenge & 4.6808 & 1.29030 & 4.5138 & 1.28407 & .16695 & .140 & 1.479 \\ \hline
\end{tabular}
\caption{Casual Players of \hp{} and \ingress{} Traits and Gratifications}
\label{tab:HPWU-ingress-casual}
\end{table}

\begin{table}[htpb]
\begin{tabular}{|r|l|l|l|l|l|l|l|}
\hline
\multicolumn{1}{|c|}{\textbf{Hardcore Players}} & \multicolumn{2}{c|}{\textbf{HPWU (288)}} & \multicolumn{2}{c|}{\textbf{\ingress{} (105)}} & \multicolumn{1}{c|}{\multirow{2}{*}{\textbf{Difference}}} & \multicolumn{1}{c|}{\multirow{2}{*}{\textbf{Sig (p)}}} & \multicolumn{1}{c|}{\multirow{2}{*}{\textbf{t}}} \\ \cline{1-5}
\multicolumn{1}{|c|}{\textbf{HPWU-\ingress{}}} & \multicolumn{1}{c|}{\textbf{Mean}} & \multicolumn{1}{c|}{\textbf{SD}} & \multicolumn{1}{c|}{\textbf{Mean}} & \multicolumn{1}{c|}{\textbf{SD}} & \multicolumn{1}{c|}{} & \multicolumn{1}{c|}{} & \multicolumn{1}{c|}{} \\ \hline
\multicolumn{1}{|l|}{\textbf{Traits}} &  &  &  &  &  &  &  \\ \hline
Aesthetic Orientation & 2.1278 & .81488 & 2.5571 & 1.07208 & -.42937 & .000 & -3.730 \\ \hline
Immersion Orientation & 2.3681 & 1.29392 & 3.3486 & 1.57701 & -.98052 & .000 & -5.709 \\ \hline
Goal Orientation & 2.3826 & 1.10462 & 2.9295 & 1.12146 & -.54688 & .000 & -4.325 \\ \hline
Social Orientation & 3.1750 & 1.57522 & 2.8229 & 1.44896 & .35214 & .046 & 2.002 \\ \hline
Action Orientation & 2.8924 & 1.24836 & 2.7352 & 1.15575 & .15712 & .261 & 1.126 \\ \hline
\multicolumn{1}{|l|}{\textbf{Gratification}} &  &  &  &  &  &  &  \\ \hline
Socialization & 4.4983 & 1.47386 & 3.0429 & 1.20144 & 1.45541 & .000 & 9.975 \\ \hline
Enjoyment & 2.3498 & 1.07403 & 2.1460 & .99992 & .20379 & .091 & 1.695 \\ \hline
Trendiness & 5.9514 & 1.28119 & 5.8603 & 1.36640 & .09107 & .541 & .612 \\ \hline
Competition & 2.1847 & .94918 & 1.8619 & .70368 & .32282 & .000 & 3.645 \\ \hline
Challenge & 3.8194 & 1.33178 & 3.8508 & 1.32367 & -.03135 & .834 & -.209 \\ \hline
\end{tabular}
\caption{Hardcore Players of \hp{} and \ingress{} Traits and Gratifications}
\label{tab:HPWU-ingress-hardcore}
\end{table}

\section{Discussion}
Our results suggest the distinctions between the core loops of the studied games differ significantly, and as such, an understanding of the three games is necessary to contextualize the results \cite{holm2019player}. 
We describe the games and their loops in Section \ref{core-loops}. To repeat, game mechanics in games are “rules” of the game; for the sake of this discussion, this definition is too granular \cite{brathwaite2009challenges}.
Most of the momentary mechanics of the three games are identical, chiefly due to their nature as LBG. 
For example, all three games feature an exploration mechanic realized through actually walking to physical locations. 
Game dynamics represent the emergent patterns of play once mechanics are set in motion by players  \cite{brathwaite2009challenges}. 
Tondello et al. \cite{tondello2017framework} argued in favor of this level of granularity for discussing games as it contextualizes game mechanics for individual games. 
These game dynamics may then be arranged in sequences that constitute the core loop of the game \cite{kiiski2019,fan2019}. 

In this section, we discuss what the results of our Traits and Gratifications analysis mean.
Our study presents the largest survey of  the traits and gratifications of casual and hardcore players in the LBG  \pogo{},  \hp{}, and  \ingress{} to date.

In terms of age, players of  \hp{} and \pogo{} tended to be under 36, while players of  \ingress{} were older. 
This distinction may be due to player bases naturally being pulled from existing fandoms, resulting in their demographics overlaping. 
The player base of the three games is relatively homogenous, with most studied players being college-educated or better and male. 
Interestingly, the latest of the three games, \hp{}, has a nearly 50/50 male to female ratio, suggesting it has gameplay elements that result in a broader appeal in terms of male and female genders. 
While fewer studies exist for \hp{} and \ingress{}, these results appear to align with other demographic studies of \pogo{}  \cite{hamari2019uses}. 
Our results also indicate that approximately 2\% of the surveyed players identify as non-binary.

Importantly, hardcore players play more, reinforcing previous assertions on hardcore player behaviors \cite{ip2005segmentation}.  
In all three LBG, players tended to be casual players on the whole. 
The player base of \hp{} was the most casual, with 72.8\% self-identifying as casual players; \pogo{} immediately followed at 67.6\%, and \ingress{} was the least casual at 60\%.
Broadly these numbers may indicate that players of these LBG tend to be more casual than not.
This may be an artifact of LBG recontextualizing urban spaces as playful ones, blurring the boundaries of the virtual game world and physical world  \cite{de2009playful}. 

Assuming hardcore players are players with a generally higher capacity for imaginative play, one may postulate that the opposite could be true for casual players \cite{bateman2011player}. 
One may argue that the superposition of the game over the real world reduces the requirement for imaginative play, lowering the barrier for entry to non-hardcore players. 
This sentiment may not necessarily explain the distinction between the three games, as they all leverage the pervasive nature of LBG. 
The hardcore casual split may more likely be the result of differences in the core loops and gameplay dynamics of the three games.

In the case of the aesthetic trait, players both hardcore and casual had a tendency to appreciate aesthetic experiences the same in most of the games, \pogo{} being the only outlier. 
In \pogo{}, casual players were more likely to have a higher affinity for the aesthetic trait than their hardcore counterparts. 
As the aesthetic orientation targets the player’s preference towards exploring the game world and experiencing the audio-visual components of the game, this trait dimension may be less related to the actual gameplay itself and more towards the overall experience \cite{tondello2019don}. 
As there’s no obvious evidence to suggest that a higher preference for aesthetic experiences as a hallmark of casual players within this dataset, the question should be asked: 
why are \pogo{} casual players more aligned to this dimension? 
To better understand any possible explanations the three games should be compared for how well they cater to aesthetic preferences.

\ingress{} design-wise, and survey-wise, is the least aesthetically oriented game. 
There are minimal options to adjust the player appearance and the UI is deliberately designed to be more utilitarian. 
A recent update to the game \textit{Ingress Prime} addresses aesthetic concerns, but at the time of the survey, it hadn’t been released \cite{Kumparak2018}. 
Both \pogo{} and \hp{} were roughly the same on average in terms of players’ aesthetic oriented traits. 
The user interface and aesthetics of these two games are generally more advanced than the first iteration of \ingress{}, although all three more or less share the same aesthetic means of world exploration as they are LBG. 
As such the subjectively more appealing user interface may explain why more players with a higher preference for aesthetics play \pogo{} and \hp{}. 

There is a key aesthetic difference between \pogo{} and \hp{}, however, players may customize their avatar in the game world in \pogo{}. 
As such, a possible weakness in the model for modeling aesthetics for games that do not have character customization may be present. 
The notion presented by Kallio in their InSoGa gaming heuristic that suggests that the game itself is a major contributor to player mentalities is supported by this observation \cite{kallio2011least}.  
Taken a step further players with particular trait mappings and gratifications may self select to games that more naturally align with their preferences.

A self selection effect may be present in player affinity for immersion. 
In all three games, players seemed to have the same affinity for immersion orientation regardless of their classification as casual or hardcore. 
Players of all three games had a preference for immersion, although to varying degrees. 
It’s reasonable to assume that in the context of LBG, affinity for immersion is a significant trait of players of the games, agreeing with the notion that LBG appropriate the real world to playful ends as suggested by Hamari et al. \cite{hamari2019uses}. 
Players of LBG likely have an affinity for immersion, as the playful behaviors of such games are generally pervasive. 
A player with a lower immersion affinity may be put off by such game dynamics, as it is not easy to play a game like \pogo{} without being immersed in the world of the game: the real world with virtual trappings. 
Our findings indicate that in order of preference for immersion, the three surveyed games ranked as such: \pogo{}, \hp{}, then \ingress{}. 
Interestingly, hardcore players of \pogo{} and \hp{} did not differ significantly in immersion preference. 

However, casual players of \pogo{} had a significantly higher affinity for immersion.
Avatar customization in \pogo{} may act as a personal stake that attracts players who have a higher affinity for immersion in the casual sphere to \pogo{}, once again reinforcing the InSoGa games heuristic assertations \cite{kallio2011least}. 
Both \pogo{} and \hp{} players had a higher immersion affinity than players of \ingress{}. 
This distinction may be down to an issue of licensing, as unlike \pogo{} and \hp{}, \ingress{} is a unique intellectual property. 
Both the Pok{\'e}mon and Harry Potter franchises are multimedia juggernauts with large existing fanbases \cite{parkin2013,kean2018}. 
Due to the high popularity of these franchises, there might be a higher population of potential players who already have an affinity for immersion, because of their fandom for the franchise. 
Ingress, on the other hand, is a newer franchise; players with immersive affinities may be players of other games or experiences which overlap with existing preferences. 
Studies of the impact of nostalgia on \pogo{} suggest that childhood brand nostalgia has a significant impact on player habit \cite{harborth2019nostalgic}. 
While not directly immersion, habit represents an analog of the immersion trait, particularly how well a game embeds a player into its story. 

As a group, hardcore players were more likely to have an affinity for the goal trait. 
This result may suggest that the typical hardcore player of LBG is more likely to play for the sake of clearing objectives. 
Such players may be more attracted to games with well-defined objectives. 
Interestingly, at first glance, this falls more in line with the original hypothesis presented in the DGD1 which suggested hardcore/casual dichotomy was how willing the player was to work to obtain victory \cite{de2009playful}. 
This manner of thinking, however, does not really answer the question of why hardcore players enjoy goals. 
If the definition arrived at by the DGD1 is used, in which hardcore players are considered “gamer hobbyists” \cite{bateman2011player}, as a perceptual lens a possible explanation can be hypothesized. 
If individual goals were considered contextualization of the core loop, each goal could be thought of as an individual gameplay session. 
While each goal should not be considered as a separate game, they do provide a framework for explorative play, wherein a player may attempt different play styles to achieve the goal.

Gamer hobbyists are marked by the range and diversity of games played, as a result, the breadth of possible experiences presented by objectives may allow hardcore players to scratch the psychological itch that causes them to pursue such a breadth of games. 
In terms of individual games, \hp{} players had the highest goal orientation. 
This is likely due to the fact that complete the foundables registry is codified as core dynamic, once again reinforcing Kallio’s assertion \cite{kallio2011least}. 
It should be noted that historically the Pok{\'e}mon franchise is notorious for its emphasis on completing the pok{\'e}dex, \pogo{} does not enforce this nearly to the level that earlier marketing had. 
Instead, the raising of a team for battles is more emphasized by the core loop, while this is still a goal completion is less defined than completing a checklist. 
However, the existence of the “Gotta Catch ‘em All” mindset appears to have skewed the casual players of \pogo{} significantly more likely to enjoy goals than the casual players of \ingress{}. 
Interestingly, while \ingress{} players had a lower affinity for goals than \hp{}, the hardcore of \ingress{} and \pogo{} didn’t significantly differ. 
This seems to support the idea that hardcore affinity for goal-based play is higher in the absence of high goal focus in the core game loop.

Much like the goal trait, an affinity for the action trait appears to serve as a predictor for the player’s status as a hardcore player. 
In all three games, the hardcore players were significantly more likely to have an affinity for challenges in the games they played. 
Once again, this aligns with the notion that hardcore players are more interested in the intensity of gaming sessions \cite{ip2005segmentation}. 
This preference for intensity suggests that being a gamer hobbyist  \cite{bateman2011player} should not be considered the sole determinate of an individual’s status as a hardcore player. 
It would be more illustrative to see if these affinities are consistent across multiple game types, as it is possible that action and goal orientations are a quirk of the hardcore LBG player. 
Interestingly, the casual player of \pogo{} is more likely to have an affinity for action than their \hp{} counterpart; likewise, the \hp{} casual has a higher affinity than the \ingress{}. This may be due to the framing of challenges in the three games.
The challenge/action of \ingress{} lies in the creation of fields dynamic, which, as established in the assessment of the social trait, is an elder game dynamic not necessarily reached by the casual players. 
In \pogo{} competition is derived from defeating other players in battle and capturing gyms. 

Gym battles tangibly pit player against player adversarially leveraging human territoriality, as different teams of players are pitted against one another for limited spatial resources.
In a study of LBG, players expressed a drive to capture and protect their home territory (a concept non-existent in the study game), implying a tendency for LBG to evoke territoriality  \cite{papangelis2017conquering,papangelis_get_2017,papangelis_unfolding_2017,chamb}. 
Interestingly, despite being available at level one the territoriality dynamics of \ingress{} are somewhat less accessible than \pogo{}. 
Players challenging gyms are able to actually see other players attacking the gym in \pogo{}, while in \ingress{} other players from your team aren’t easily visible in the game UI.
Additionally, while \pogo{} gyms tend to have high level teams, they also leverage skillful play to allow challengers of lower levels to take down gyms on their own. 
In the case of \ingress{} portals captured by higher level players are not as easy for new players to take down, requiring a heavier focus on socialization. 
As casuals are apparently less affine to social dynamics this seems to be a possible explanation for the disparity between the casuals of \pogo{} and \ingress{}. 
Meanwhile \hp{} offers minimal competition, instead focusing on the challenge of capture the foundables and participating in co-op battles. 
The capturing of foundables likely is more emphasized in the minds of the players as a goal-based dynamic, while the co-op battles lack the inherent rivalry present in a pok{\'e}mon battle. 

Of the studied games, only \ingress{} had a significant difference between hardcore and casual players regarding the social trait. 
Hardcore players tended to be significantly more social than casual. 
In \ingress{}, the create fields dynamic intrinsically requires social interaction for effective play. 
A solo player will struggle to maintain three portals (the requisite number to generate a field), and gather enough resources to connect those portals to generate a field. 
While a part of the core loop, it can also be considered a part of the elder game of \ingress{}, meaning a player will need to devote more time to the game to reach this gameplay dynamic.
This may suggest that hardcore players are more likely to dedicate more time to games, which would support the assertion that hardcore players play longer sessions, more often  \cite{ip2005segmentation}. 
This pattern does not emerge in the other games; however, as social aspects of the two other games are more easily accessible from the start of a player’s experience with the game. 

When comparing the games, \pogo{} players were significantly more likely to enjoy social interaction than their \hp{} counterparts. 
As \pogo{} emphasizes taking on gyms and raids in collaborative battles as a core dynamic, this makes sense; however, \hp{} also has a core dynamic revolving around co-op play.
It might be possible that the emphasis on the collectible aspects of the gameplay overshadows the social elements, and as such, players more focused on individual goals are more likely to enjoy \hp{}. 
\hp{} casual players were more likely to enjoy socializing with other players than their \ingress{} counterparts. 
Likewise, \ingress{} hardcore players are more likely to have an affinity for the social trait than \hp{} hardcore players. 
This may reinforce the assertion that hardcore players of \ingress{} are more driven by the social dynamics enforced by the create fields dynamic, and casual players simply don’t engage in that behavior as much. 
It’s also possible that the Harry Potter franchise is a higher motivator for socialization in the casual set of players. 

Traits target general player attributes, while gratification questions are more targeted towards the manner in which a player experiences a specific game. 
When surveyed for socialization gratification, hardcore players were generally more gratified by the game-related social activities. 
This reinforces the findings in assessing the social trait where hardcore players were typically more attuned. 
Positive gratification from socialization agrees with findings in which socialization positively influences continued play for users of \pogo{}  \cite{alha2019people}. Additionally, this further supports the view that hardcore players socialize over games, as presented by Bateman and Nacke \cite{alha2019people}. 
Interestingly, in \hp{} casual players were actually neutral to the socialization gratification, perhaps suggesting \hp{}’s core-loop is less suited to social play, at least for casual players. 
The emphasis on goals and collecting may once again focus players of \hp{} on a more solo experience. 

These findings also suggest that players are generally less satisfied by social interaction than they expect themselves to be. 
In all three cases, players were less gratified by socialization than they had assessed in the trait section of the survey. 
This may suggest that players aren’t necessarily good at predicting gameplay they enjoy.
Tondello’s work on custom gamification supports this, wherein traits were not necessarily good indicators of game element selection  \cite{tondello2019dynamic}. 
It follows that while the players may have an affinity for a specific trait, the execution of those elements may not sufficiently gratify the players. 
Meaning this is less of a problem with the player’s ability to identify what they like and how well the game delivers on their perception of that gratification or dynamic. 
\ingress{} players are most gratified by game-related social activity, once again reinforcing the importance of social activity in the core loop of the game. 
\pogo{} was next, followed by \hp{}, which lines up with the emphasis on the social elements in the games’ core loops as described in the discussion of the social trait.

Hardcore players are generally more gratified by their enjoyment of the game. 
Questions pertaining to enjoyment were framed around how the player perceived the gameplay itself, not side benefits of the games. 
The high gratification for hardcore players suggests that the hardcore players of LBG enjoy games for games’ sake, a defining trait of gamer hobbyists, reinforcing Bateman and Nacke’s claim \cite{bateman2011player}. 
This is not to say that casual players are incapable of enjoying games for games’ sake, in fact, casual gratification was fairly high in all three games. 
\ingress{} players seemed to be the most gratified by their enjoyment of the games; however, the only significant difference is between the casual players of \ingress{} and \hp{}.
Casual players of \hp{} are less gratified by the enjoyment of the game than casual players of \ingress{}. 
This distinction may be because \hp{} has a tight coupling to the Harry Potter franchise, meaning players may be more gratified by feelings of nostalgia or appreciation of the franchise in general. 
This distinction appears to partially support the finding that childhood brand nostalgia is positively associated with hedonic motivation, an analog for enjoyment in behavioral intention, however, \pogo{}’s lack of difference is unexplained and warrants further study  \cite{harborth2019nostalgic}.

Of the gratifications surveyed, trendiness was the least gratifying, unilaterally, aligning with Hamari’s initial findings on \pogo{} \cite{hamari2019uses}. 
Regardless of the influence of trendiness on adoption, it seems not to be a valuable gratification for playing LBG. Trends are responsible for the initial adoption of a game, but not the continued interaction between players and the game. 
For repeat play, games need to establish an affective relationship with the player  \cite{jenkins2008convergence}.
Trendiness only provides the initial dopamine hit to install the game on the player’s device; the gameplay does not influence the actual gratification. 
There was a significant relationship between the hardcore players of \pogo{} and \hp{}. 
\pogo{} hardcore players were more gratified by trendiness. Players are swayed by fads, and this discretion may be an artifact of pogo{}’s meteoric rise to prominence in 2016  \cite{fitzpatrick2018}. 
Ingress had no significant difference with any of the other games. 
This may be interpreted as the game being less linked to trendiness, as it doesn’t have the built-in franchise of the other games.

In each of the games, players receive gratification by competing with other players in the context of the game. 
Hardcore players are significantly more gratified by competition in all of the surveyed games, suggesting competition is a common gratification for hardcore players of LBG.
Gratification for competition in hardcore players aligns with the findings of the socialization gratification and social trait of this study. 
Competition represents a social dynamic in LBG, as players fight for control over physical locations, they interact with members of opposing teams in the game. 
The core loop of \ingress{} revolves around a competitive struggle between two player factions (the Resistance and the Enlightened), in which players vie for limited resources  \cite{hatfield2014}. 
The social dynamic created by this struggle supports the notion that hardcore players are gamer hobbyists  \cite{bateman2011player}, as competition provides an outlet for communication through gameplay dynamics.

Competition also represents an intensity of gameplay, which aligns with the action trait, supporting the view that hardcore players prefer intense experiences  \cite{ip2005segmentation}. 
The multifaceted nature of competition appears to support that Hardcore players are not merely gamer hobbyists or action seekers, but a combination of the two groups. 
Breaking competition down by games, \ingress{} stands supreme, with its players receiving the most gratification from competition. 
As competition is key to the core loop of \ingress{}, this comes as no surprise. 
When comparing \hp{} and \pogo{}, casual players of \pogo{} receive more gratification from competition than their \hp{} counterparts; however, the opposite is true for hardcore players of the same games. 
Suggesting that the competitive elements of \pogo{} may be better suited to casual players, while the competitive aspects of \hp{} are better suited to hardcore players. 
A more thorough investigation of the competitive dynamics of the two games may be warranted to better tailor experiences for hardcore and casual players.

Response to the challenge gratification was overall neutral or negative. 
As both the challenge and competition gratifications appear to logically model elements of the action trait, this is a fairly surprising finding. 
In each of the games, hardcore players scored neutrally on this gratification, while casual players scored negatively. 
While this generally indicates a greater affinity for challenge in hardcore players, it appears to paint a broader picture of LBG players in general when considered in terms of competition. 
These findings suggest that while players of LBG express the action trait, they are more gratified by action in a social context. 
Due to the pervasive nature of LBG, players may have a different expectation of such games, wherein socialization is more valued and enhances the gratification of gameplay dynamics. 

Our finding points to human territoriality driving gratification in LBG, supporting and strengthening the findings of Papangelis et al.  \cite{papangelis2017conquering, papangelis_get_2017, perf}. 
The findings also suggest that this sense of territoriality is experienced regardless of gender, as in all three games a similar trend was observed with regards to the dichotomy of challenge and competition. 
As previous research only observed this in males our findings indicate a more broad trend amongst players of LBG. 
As this model has only been applied in the context of LBG, further research is required to determine if this split is an aspect of LBG or gaming in general. 
It’s possible that the socialization offered by competition, on the whole, is a better gratifier for players in games in the general context. 
In comparing the games, a significant distinction was only found in casual players of \pogo{} being less gratified by challenge than casual players of \ingress{}. 
This affinity for challenges may support the notion that players of \ingress{} as a whole are skewed more towards the hardcore, assuming the assumption that hardcore players are more tolerant of challenge is valid. 
\hp{} had no significant differences with the other studied games.

As alluded to before, \pogo{} and \hp{} are just an element of two massive transmedia franchises  \cite{parkin2013, kean2018}. 
In both games, nostalgia was an apparent gratification regardless of the player’s status as casual or hardcore. 
In \hp{}, hardcore players received significantly more gratification than their casual counterparts. 
As nostalgia doesn’t exist in a vacuum, this may suggest that people with a higher nostalgia for the Harry Potter franchise may be more likely to be hardcore players of \hp{}.
\hp{} emphasizes gameplay dynamics around the franchise (foundables are items, creatures, and people from the franchise as a whole). 
Overall this reinforces the findings of Rauschnabel et al. and Hamari, which found nostalgia to be a driving force for players of \pogo{}  \cite{hamari2019uses,rauschnabel2017adoption}.  

The overall high gratification derived from nostalgia and the higher response rate for \pogo{} and \hp{} may indicate that nostalgia drives interest in LBG. 
While not measured \ingress{} is unique among the studied games in terms of nostalgia.  
As a unique intellectual property at the time of release \ingress{} has no childhood brand nostalgia experienced by its contemporaries and there was no transmedia franchise to support the gameplay. 
A recent anime\footnote{\url{https://www.netflix.com/gb/title/80992853}} and book series\footnote{\url{https://www.amazon.co.uk/Ingress-Niantic-Project-Files-Book/dp/B01BOWID4C}}represents an effort by Niantic to build a broader transmedia franchise in an effort to leverage an analog of this gratification. 
Leveraging immersion Niantic even created integrated events for the anime release in an effort to drive interaction  \cite{sherman2018,papangel-board}. 
A study on the impacts of these efforts is warranted, although an analog of the nostalgia gratification may need to be constructed.

Based on our findings it appears that in the context of LBG casuals may almost be considered a baseline player. 
No single attribute, trait or gratification, had a higher casual affinity than hardcore affinity. 
In fact casual only beat out hardcore once in the case of the affinity for aesthetics in \pogo{}. 
In contrast hardcore players were more goal, social and action oriented and received more gratification from socialization, enjoyment, competition, and challenge. 
Our findings reinforce the notion that hardcore players may be generally considered gamer hobbyists, however, they also appear to seek intense gameplay experiences \cite{ip2005segmentation,bateman200521st}.
Social dynamics on the whole appeared to increase gratification for LBG players. 
Further application of the model to a non-LBG context should occur to determine if this is a unique feature of LBG or a general trend in games as a whole. 

Interestingly, there also appears to be some discrepancy between the measurement of traits and the perceived gratifications. 
Despite clear analogs, such as the social trait and socialization gratification, gratification appears to be lower than the measurement of the player’s trait on a similar Likert scale. 
The mechanism for traits may not be indicative of gratifications, or perhaps games may not properly deliver on the gratifications measured. 
Similar trends were found to be existing in prior work by Tondello, as such further study is recommended  \cite{tondello2019dynamic}. 

Finally, gameplay dynamics and core loops appear to influence the players of their games.
Games with design elements conducive to certain traits appear to attract players with those traits (e.g. goal affine players to \hp{}). 
Gratifications also appeared to have correlation to the independent games, with variations in levels of gratification correlating to the actual dynamics of those games. 
In sum, our findings point to a collection of trends along the hardcore and casual players of the three studied game, presenting a preponderance of commonality between the three games, while also indicating discrete attributes borne from the individual game elements.

\section{Conclusion}

This study presents one of the most comprehensive explorations of the links between player traits and gratifications in LBG.
Concurrent to the exploration of LBG, this work presents the most extensive scholarly study of the player bases of \pogo{}, \hp{}, and \ingress{}. 
In particular, our work sought to determine three attributes of the communities mentioned above: 

    \begin{enumerate}
        \item The casual and hardcore players of the games.
        \item The traits  casual and hardcore players manifest in the games. 
        \item The gratifications which drive the casual and hardcore players in the games.
    \end{enumerate}

To this end, we deployed a survey with three components: 
a demographic component, 
a version of Tondello’s trait instrument  \cite{tondello2019dynamic}, and 
a version of Hamari’s gratifications instrument  \cite{hamari2019uses}. 
The surveys were deployed to /r/PokemonGo, /r/HPWU, and /r/Ingress in mid-July 2019, and the responses were collected for ten days. 
After data filtering (e.g. removing incomplete surveys or surveys with a single answer for all items) 1071, 1057, and 262 surveys were collected for \pogo{}, \hp{}, and \ingress{}, respectively.  
Exploratory and confirmatory factor analysis conducted on the data sets found the data sets to be valid, and the factor loadings to be generally valid for the survey items.


Our survey results appear to indicate distinct attributes for the casual and hardcore players of the studied LBG. 
Remember, for the purposes of this study, hardcore players are considered a grouping of players who favor more committed, higher intensity play. 
In contrast, lower intensity, less committed players are considered casual players. By situating these two types of players within our instrumentation, we filter between heavy and light types of players which affords us an ability to more finely contextualize our results. 
Hardcore players were typically more affine for the studied traits and more gratified by the gratifications presented. 
Hardcore players were usually found to be a combination of gamer hobbyists, who derived gratification from socializing over games and the games themselves, and intense players, who valued intense gaming sessions  \cite{bateman2011player,ip2005segmentation}. 
Generally, our studied players appeared to be more gratified by socialization in all of its forms, perhaps indicating a general trend in LBG players in general. 
Finally, the design of the games was generally determined to have an influence on the measured gratifications of players and the traits expressed by those who played.

The limitations of this work are as follows.
First, in Hamari’s survey, “outdoor activity” was included as a gratification \cite{hamari2019uses}. 
Through exploratory factor analysis, it was determined that outdoor activity is a part of the “socialization” dimension in all three applications of the survey. 
This is believed to be the result of our sample having different play experiences from the original sample studied by Hamari et al.  \cite{hamari2019uses}. 
Additionally, and more importantly, our participants likely interpreted the questions as part of the socialization dimension in the questionnaire as the questions are closely related to social activities (e.g. "I play \pogo{} because I can meet friends outdoors", and "I play \pogo{} because I can meet strangers outdoors"). 

As the literature and previous studies indicate that outdoor activity is an important gratification of players of LBG, this shows a significant drawback in this study of LBG players. 
Further research is required to explore outdoor activity as a gratification. 
Due to societal changes and global pandemic in 2020 gratifications as a whole may need exploration as a whole. 
Changes in socialization introduced by the global pandemic may impact the way players experience LBG as a whole. 
It is possible that certain gratifications have been emphasized (e.g. outdoor activity) or reduced (e.g. socialization) in light of this event. 
As such many of the conclusions reached by this work may need some reconsideration in a post-COVID-19 world. 
A study of changes to player gratifications and impacts on players’ interactions with the core loops of LBG are of particular interest.

Individuals in our survey self categorize as hardcore and casual. 
Self-categorization is somewhat problematic, as player actions don’t necessarily match the characterizations they give themselves. 
This particular drawback might be highlighted by the observed disconnect between traits and gratifications discussed in the discussion section, which requires further work in its own right. 
More concretely, one individual in the study categorized themselves as casual; however, it was determined based on interactions that they would be better classified as hardcore (e.g. they would go out to play at night). 
Further research is required into the casual and hardcore players of LBG and their actions. 
In future work, it is suggested to consider the casual/hardcore categorization as more of a continuum, rather than the binary presented in this work. 

This study was conducted when \hp{} had just been released, and \ingress{} had been updated to \textit{Ingress Prime}. 
These recent changes may have influenced the results and colored the responses given by players of these games. 
In further work, it is recommended that a more “longitudinal” approach be employed. 
In this approach, samples would be gathered at different frames of time in an effort to reduce temporal artifacts. 
In this line of thinking an investigation into the accessibility of mechanics is recommended as well. 
As discussed there appears to be some impact on accessibility to gameplay mechanics and gratifications. 
In this work, a hypothetical link was presented between territoriality and the competition gratification, which warrants further study.

Finally, all of the participants of this survey have been recruited through Reddit.
Usually, the most involved and dedicated players are the regulars of such forums of discussion. 
As such, the results may have been skewed more hardcore than more general recruitment.
It is recommended for future research into LBG that recruitment be performed through other avenues, particularly in-game interactions, if possible.
Overall, it is our belief that through the usage of the above-outlined methods and exploratory studies traits and gratifications may be leveraged to acquire a more full understanding of the players of LBG.

\bibliographystyle{ACM-Reference-Format}
\bibliography{main}

\appendix

\section{Survey}

\subsection{Section 1 - Demographics and playing-related factors}

In this section we will ask you a few questions to find out more about you.

Gender
\begin{itemize}
    \item Male
    \item Female
    \item Non-Binary
    
\end{itemize}

Age 
\begin{itemize}
    \item Under 15 years
    \item 15 to 20 years
    \item 26 to 30 years
    \item 31-35 years
    \item 36-40 years
    \item 41-45 years
    \item 46-50 years
    \item Over 51 years
\end{itemize}

Occupation
\begin{itemize}
    \item Working full time
    \item Working part time
    \item Unemployed
    \item Retired/Pensioner
    \item Student
\end{itemize}

Country of residence
\begin{itemize}
    \item List of countries
\end{itemize}

Education
\begin{itemize}
    \item University degree
    \item College degree
    \item Vocational degree
    \item High school
    \item Others
\end{itemize}

I typically play GAMENAME:
\begin{itemize}
    \item Many times every day
    \item Once a day
    \item A few times a week
    \item Once a week
    \item A couple of times a month
    \item Rarely
\end{itemize}

How long each game session usually lasts?
\begin{itemize}
    \item Less than 15 minutes
    \item 16-30 minutes
    \item 31-45 minutes
    \item 46-60 minutes
    \item 1-2 hours
    \item 2-3 hours
    \item 3-4 hours
    \item 4-5 hours
    \item 5-6 hours
    \item 6 hours or more
\end{itemize}

Would you consider yourself a casual or a hardcore GAMENAME gamer?
\begin{itemize}
    \item Casual
    \item Hardcore
\end{itemize}


\subsection{Section 2 – Player traits}

\textit{Notes: 7-point Likert Scale for all items (Strongly agree to Strongly disagree)}

In the following section we will ask few questions that will help us understand what kind of gamer you are. In contrast with the rest of the questionnaire, the questions in this section are general and not specific to Harry Potter Wizard Unite.

Aesthetic orientation
\begin{enumerate}
    \item I like games which make me feel like I am actually in a different place. 
    \item I like games with detailed worlds or universes to explore. 
    \item I like to spend some time exploring the game world. 
    \item I like to customize how my character looks in a game. 
    \item I often feel in awe with the landscapes or other game imagery. 
\end{enumerate}

Narrative orientation
\begin{enumerate}
    \item I usually skip the story portions or the cutscenes when I am playing. (R)
    \item Story is not important to me when I play games.  (R)
    \item I enjoy complex narratives in a game. 
    \item I like games that pull me in with their story. 
    \item I feel like storytelling often gets in the way of actually playing the game. (R)
\end{enumerate}

Goal orientation
\begin{enumerate}
    \item I like to complete all the tasks and objectives in a game. 
    \item I usually do not care if I do not complete all optional parts of a game (R)
    \item I like completing games 100
    \item I feel stressed if I do not complete all the tasks in a game. 
    \item I like finishing quests. 
\end{enumerate}

Social orientation
\begin{enumerate}
    \item I like to play online with other players. 
    \item I like to interact with other people in a game. 
    \item I don’t like playing with other people.  (R)
    \item I like games that let me play in guilds or teams. 
    \item I often prefer to play games alone. (R) 
\end{enumerate}

Challenge Orientation
\begin{enumerate}
    \item I like it when goals are hard to achieve in games. 
    \item I enjoy highly difficult challenges in games. 
    \item I like it when games challenge me. 
    \item I usually play games at the highest difficulty setting. 
    \item I like it when progression in a game demands skill. 
\end{enumerate}

\subsection{Section 3 - Gratifications}

\textit{Notes: 7-point Likert Scale for all items (Strongly agree to Strongly disagree) except nostalgia which is Yes/No.}

In this section we ask you questions that will help us understand how and why you play GAMENAME. 

Challenge
\begin{enumerate}
    \item I feel proud when I master an aspect of the game 
    \item It feels rewarding to get to the next level 
    \item I feel excited when I return foundables to their proper location
    \item I feel excited when I win a battle 
    \item I enjoy finding new and creative ways to work through the game
\end{enumerate}

Competition
\begin{enumerate}
    \item I like to prove that I am one of the best players 
    \item I get upset when others do better than me 
    \item I get upset when I am unable to earn enough points 
    \item It is important to me to be one of the skilled persons playing the game
\end{enumerate}

Enjoyment
\begin{enumerate}
    \item I play GAMENAME because it is exciting 
    \item I play GAMENAME because it is entertaining 
    \item I play GAMENAME because it is fun
    \item I play GAMENAME because it is a good pastime 
    \item I play GAMENAME because it is a habit 
    \item I play GAMENAME because it occupies my free time 
\end{enumerate}

Trendiness
\begin{enumerate}
    \item GAMENAME enables me to look trendy
    \item GAMENAME enables me to look cool
    \item GAMENAME enables me to look stylish 
\end{enumerate}

Socializing
\begin{enumerate}
    \item GAMENAME enables me to maintain friendships 
    \item GAMENAME enables me to improve relationships 
    \item GAMENAME enables me to make new friends
    \item I like to play GAMENAME because my friends play the game 
    \item GAMENAME enables me to participate in relevant discussions
    \item GAMENAME enables me to be part of a group
\end{enumerate}

Outdoor activity
\begin{enumerate}
    \item I play GAMENAME because it motivates me to go out 
    \item I play GAMENAME because it motivates me to explore new places 
    \item I play GAMENAME because I can meet friends outdoors 
    \item I play GAMENAME because I can meet strangers outdoors
\end{enumerate}

Nostalgia (Yes/No for this section; no 7-point Likert Scale)\footnote{These questions are not included in the \ingress{} survey.}
\begin{enumerate}
    \item I used to read Harry Potter books 
    \item I used to play Harry Potter PC/Console games
    \item I used to watch Harry Potter cartoons/anime series/movies 
    \item I used to collect Harry Potter merchandise (e.g. toys, stickers, trading cards, etc.) 
    \item I have been a fan of Harry Potter even before the launch of GAMENAME
\end{enumerate}

\end{document}